\documentclass[12pt]{article}

\usepackage{axodraw}
\usepackage{amssymb}
\usepackage{epsfig}

\usepackage[normalem]{ulem}
\usepackage{xcolor}
\usepackage{physics}
\usepackage[symbol]{footmisc}
\renewcommand{\thefootnote}{\fnsymbol{footnote}}

\catcode`@=11
\def\citer{\@ifnextchar [{\@tempswatrue\@citexr}{\@tempswafalse\@citexr[]}}
 
\def\@citexr[#1]#2{\if@filesw\immediate\write\@auxout{\string\citation{#2}}\fi
  \def\@citea{}\@cite{\@for\@citeb:=#2\do
    {\@citea\def\@citea{--\penalty\@m}\@ifundefined
       {b@\@citeb}{{\bf ?}\@warning
       {Citation `\@citeb' on page \thepage \space undefined}}%
\hbox{\csname b@\@citeb\endcsname}}}{#1}}
\catcode`@=12
 
\reversemarginpar

\newcommand{\lsim}{\raisebox{-0.13cm}{~\shortstack{$<$ \\[-0.07cm] $\sim$}}~}

\newcommand{\slsh}[1]{\!\!\not\! {#1}}

\newcommand{\tgb}{{\rm tg} \beta}

\newcommand{\sgl}{{\tilde{g}}}
\newcommand{\stopx}{{\tilde{t}}}
\newcommand{\sbottom}{{\tilde{b}}}

\newcommand{\sq}{{\tilde{Q}}}

\newcommand{\LF}[1]{#1}

\newcommand{\LFcomment}[1]{}

\begin{document}

\renewcommand{\thefootnote}{\fnsymbol{footnote} }

\vskip-1.0cm

\begin{flushright}
CERN--TH--2022--100 \\
KA--TP--16--22 \\
PSI--PR--22--16
\end{flushright}

\begin{center}
{\large\sc Pseudoscalar MSSM Higgs Production at NLO SUSY--QCD}
\end{center}

\begin{center}
Emanuele Bagnaschi$^1$, Lukas Fritz$^{2,3}$, Stefan Liebler$^4$\footnote[3]{Former academic affiliation},
Margarete M\"uhlleitner$^4$, Thanh Tien Dat Nguyen$^4$ and Michael
Spira$^2$
\end{center}

\begin{center}
{\it \small
$^1$ Theoretical Physics Department, CERN, CH--1211 Geneva 23, Switzerland \\
$^2$ Paul Scherrer Institut, CH--5232 Villigen PSI, Switzerland \\
$^3$ Institut f\"ur Theoretische Physik, Zurich University, CH--8057
Zurich, Switzerland \\
$^4$ Institute for Theoretical Physics, Karlsruhe Institute of
Technology, D--76128 Karlsruhe, Germany}
\end{center}

\begin{abstract}
\noindent
One of the most important mechanisms at the Large
  Hadron Collider (LHC) for the production of the
  pseudoscalar Higgs boson of the Minimal Supersymmetric Standard
  Model (MSSM) is the loop-induced gluon fusion process $gg\to A$. The
higher-order QCD corrections have been obtained a long time ago and
turned out to be large. However, the genuine supersymmetric (SUSY--)QCD corrections have been obtained only in the limit of large SUSY particle masses so far. We describe our calculation of the next-to-leading-order (NLO) SUSY--QCD results with full mass dependence and present numerical results for a few representative benchmark points. We also address the treatment of the effective top and bottom Yukawa couplings, in the case of heavy SUSY particles, in terms of effective low-energy theories where the heavy degrees of freedom have been decoupled. Furthermore, we include a discussion of the relation between the SUSY--QCD corrections that we have computed  and the Adler--Bardeen theorem for the axial anomaly. In addition, we apply our results to the gluonic and photonic pseudoscalar Higgs decays
$A\to gg,\gamma\gamma$ at NLO.
\end{abstract}

\def\thefootnote{\arabic{footnote}}
\setcounter{footnote}{0}

\section{Introduction}
The discovery of a Standard-Model-like Higgs boson at the LHC \cite{1}
completed the Standard Model (SM) of electroweak and strong
interactions. The existence of the Higgs boson \cite{2} is inherently
related to the mechanism of spontaneous symmetry breaking while
preserving the full gauge symmetry and the renormalizability of the SM
\cite{3}. The measured Higgs boson mass of $(125.09 \pm 0.24)$ GeV \cite{4}
ranks at the weak scale. The existence of the Higgs boson allows the SM
particles to be weakly interacting up to high-energy scales \cite{5}.
This, however, is only possible for particular Higgs-boson couplings to
all other particles, so that the knowledge of the Higgs-boson mass fixes
all its properties uniquely. The massive gauge bosons and fermions
acquire mass through their interaction with the Higgs field that
develops a vacuum expectation value in its ground state. The minimal
model requires the introduction of one isospin Higgs doublet and leads
after spontaneous symmetry breaking to the existence of one scalar Higgs
boson. The SM itself, however, leaves several fundamental questions open
as e.g.~the nature of Dark Matter, the baryon asymmetry of the universe
or the stability of the electroweak against the Planck or grand
unification scale. If the SM is extended to a Grand Unified Theory (GUT)
scale, radiative corrections to the Higgs-boson mass tend to push it
towards the GUT scale, if the Higgs boson couples to particles of that
mass order. In order to obtain a Higgs mass at the electroweak scale the
Higgs-mass counterterm has to be fine-tuned to cancel these large
corrections thus establishing an unnatural situation that asks for a
solution. This is known as the hierarchy problem \cite{6}. These open
questions call for extensions of the minimal model. To increase the
experimental sensitivity to effects beyond the SM (BSM), the SM and BSM
parts of measured relevant observables need to be known as precisely as
possible in order to allow for a reliable interpretation of potential
deviations and effects beyond the SM.

The open problems of the SM motivate extensions of the minimal model
which cover e.g.~the Two-Higgs-Doublet model (2HDM) \cite{2hdm} or the
minimal supersymmetric extension (MSSM) \cite{mssm,haberkane} as
prominent and highly motivated examples. Supersymmetric extensions of
the SM provide a solution to the hierarchy problem if the supersymmetric
particle masses rank at scales up to a few TeV \cite{10}. Supersymmetry
relates fermionic and bosonic degrees of freedom and thus links internal
and external symmetries. The MSSM, if embedded in a Grand
Unified Theory, predicts a value of the Weinberg angle in excellent
agreement with experimental measurements of electroweak precision
observables \cite{11}. Moreover, it contains a Dark Matter candidate if
R-parity is conserved \cite{12} and allows for generating electroweak
symmetry breaking radiatively, since the top mass ranks in the proper
region for that mechanism to work \cite{13}. The MSSM introduces two isospin Higgs doublets due to
the analyticity of the superpotential, requiring two different doublets
for the generation of the up- and down-type fermion masses and the
anomaly-freedom with respect to the gauge symmetries \cite{14}, since
the higgsino states as the supersymmetric partners of the Higgs bosons
contribute to the Adler-Bell-Jackiw anomaly \cite{ABJ}. Due to this, the
MSSM Higgs sector is a 2HDM of type II at leading order (LO). There are a light ($h$) and
heavy ($H$) scalar, a pseudoscalar ($A$) and two charged ($H^\pm$)
states as the corresponding mass eigenstates. Since the
self-interactions of the Higgs fields, as defined by the corresponding
Higgs potential, are entirely fixed by the electroweak gauge couplings,
this induces an upper bound on the light scalar Higgs mass that has to
be smaller than the $Z$-boson mass $M_Z$ at LO. However,
radiative corrections, which are dominated by top-quark-induced
contributions, strongly increase this upper bound to about 130 GeV in
general \cite{15}. The Higgs sector is uniquely fixed at LO by the value
of the pseudoscalar mass $M_A$ and the parameter $\tgb$, defined as the
ratio of the two vacuum expectation values of the scalar Higgs fields.

In this work, we will describe the calculation of the full SUSY--QCD
corrections at NLO to pseudoscalar Higgs production via the gluon-fusion
mechanism $gg\to A$. This process belongs to the dominant MSSM Higgs-boson production processes at the LHC and thus contributes to the present bounds on the so far negative searches for the heavy MSSM Higgs bosons at the LHC. In order to make the predictions for this process reliable, the full NLO corrections within SUSY--QCD have to be computed. The paper is organized as follows. In Section
\ref{sc:glufus}, we will summarize the present status of the gluon-fusion
cross section. In Section \ref{sc:adec}, we briefly discuss pseudoscalar
Higgs decays to gluons and photons. In Section \ref{sc:squark} we will
describe our implementation of the stop and sbottom sector followed by
the detailed description of our NLO calculation in Section \ref{sc:nlo}.
In the latter we also include a discussion of effective Yukawa couplings 
and the relation of the considered process to the Adler--Bardeen theorem \cite{adlerbardeen}. In Section \ref{sc:results}, we discuss numerical results for a few
representative benchmark points. We close the paper with our conclusions in Section \ref{sc:conclusions}.

\section{Gluon Fusion} \label{sc:glufus}
The dominant channels for pseudoscalar production at a hadron
  collider are given by
gluon fusion, $gg\to A$, and production in association with
bottom quarks, $q\bar q,gg\to Ab\bar b$, with their
relative importance depending on the value of $\tgb$.
For large $\tgb$, $Ab\bar b$ production dominates, with the gluon fusion contribution
amounting to up to about 30\% close to the present exclusion bounds, depending 
on the region in the $M_A - \tgb$ plane \cite{hprod, bench}.

\subsection{Leading Order}
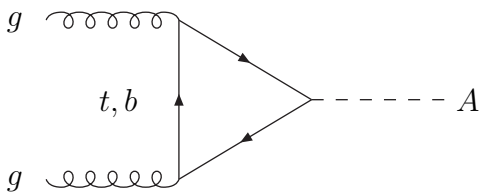
\begin{figure}[hbt]
\begin{center}
\setlength{\unitlength}{1pt}
\begin{picture}(180,100)(0,0)
\Gluon(0,20)(50,20){-3}{5}
\Gluon(0,80)(50,80){3}{5}
\ArrowLine(50,20)(50,80)
\ArrowLine(50,80)(100,50)
\ArrowLine(100,50)(50,20)
\DashLine(100,50)(150,50){5}
\put(155,46){$A$}
\put(20,46){$t,b$}
\put(-15,18){$g$}
\put(-15,78){$g$}
\end{picture}  \\
\setlength{\unitlength}{1pt}
\caption{\label{fg:ggalodia} \it Typical diagram contributing to $gg\to
A$ at lowest order.}
\end{center}
\end{figure}

The gluon-fusion mechanism \cite{glufus}
\begin{displaymath}
pp \to gg \to A
\end{displaymath}
dominates the pseudoscalar MSSM Higgs boson production at the LHC in the
phenomenologically relevant Higgs mass ranges for small and moderate
values of $\mbox{tg$\beta$}$. Only for large $\mbox{tg$\beta$}$ the
associated $A b\bar b$ production channel develops a larger cross section
due to the enhanced Higgs couplings to bottom quarks \cite{htt}. The
gluon coupling to pseudoscalar Higgs bosons in the MSSM is built up by
loops involving top and bottom quarks, see Fig.~\ref{fg:ggalodia}.
The partonic cross section is given at lowest order by \cite{hunter,higgsqcd}:
\begin{eqnarray}
\hat\sigma^A_{LO} (gg\to A) & = & \sigma^A_0 \delta
(1 - z) \label{eq:mssmggalo} \nonumber \\
\sigma^A_0 & = & \frac{G_{F}\alpha_{s}^{2}(\mu_R)}{128 \sqrt{2}\pi} \
\left| \sum_{Q} g_Q^A A_Q^A (\tau_{Q}) \right|^{2} \;,
\label{eq:locxn}
\end{eqnarray}
where $G_F$ denotes the Fermi constant, $\alpha_s$ the
  strong coupling, and $\mu_R$ the renormalization scale. The scaling variables are defined as $z=M_A^2/\hat s$, $\tau_Q
=4M_Q^2/M_A^2~~(Q=t,b)$, and $\hat{s}$ denotes the partonic c.m.~energy
squared. The amplitudes $A_Q^A (\tau_Q)$ are obtained as
\begin{eqnarray}
A_Q^A (\tau) & = & \tau f(\tau) \nonumber \\
f(\tau) & = & \left\{ \begin{array}{ll}
\displaystyle \arcsin^2 \frac{1}{\sqrt{\tau}} & \tau \ge 1 \\
\displaystyle - \frac{1}{4} \left[ \log \frac{1+\sqrt{1-\tau}}
{1-\sqrt{1-\tau}} - i\pi \right]^2 & \tau < 1
\end{array} \right.
\label{eq:loff}
\end{eqnarray}
and the MSSM coupling factors $g_Q^A$ are determined as $g_t^A =
1/\tgb$, $g_b^A = \tgb$.  In the narrow-width approximation the hadronic
cross section is given by
\begin{equation}
\sigma_{LO}(pp\to A) = \sigma^A_0 \tau_A \frac{d{\cal L}^{gg}}
{d\tau_A}
\end{equation}
with the scaling variable $\tau_A = M^2_A/s$, where $s$ specifies the total hadronic c.m.~energy squared, and the gluon luminosity
\begin{equation}
\frac{d{\cal L}^{gg}}{d\tau} = \int_\tau^1 \frac{dx}{x}~g(x,\mu_F^2)
g(\tau /x,\mu_F^2)
\label{eq:gglum}
\end{equation}
at the factorization scale $\mu_F$. For
small $\tgb$ the top-loop contribution is dominant, while for large
values of $\tgb$ the bottom-quark contribution is strongly enhanced.

\subsection{QCD Corrections}
The full two-loop QCD corrections to the gluon-fusion cross
section were calculated in the past
\cite{higgsqcd,gghsusy,hgaga1}. In complete
analogy to the SM case, they consist of virtual two-loop corrections to
the basic $gg\to A$ process and real one-loop corrections due to the
associated production of the pseudoscalar Higgs boson with massless
quarks and gluons. The final result for the hadronic cross section at
NLO can be decomposed as
\begin{equation}
\sigma(pp \rightarrow A +X) = \sigma^A_{0} \left[ 1+ C^A
\frac{\alpha_{s}}{\pi} \right] \tau_A \frac{d{\cal
L}^{gg}}{d\tau_A} +
\Delta\sigma^A_{gg} + \Delta\sigma^A_{gq} +
\Delta\sigma^A_{q\bar{q}} \;.
\label{eq:mssmgghqcd5}
\end{equation}
The analytical expressions for arbitrary Higgs boson and quark masses at
NLO are rather involved \cite{higgsqcd,hgaga1}. As in the SM case, the
quark-loop masses have been identified with the pole mass $m_Q$
$(Q=t,b)$, while the QCD coupling and the parton distribution
functions (PDFs) of the proton are treated in
the $\overline{\rm MS}$ scheme with five active flavours. The axial
$\gamma_5$ coupling can be regularized in the 't~Hooft--Veltman scheme
\cite{thoovel} or its extension by Larin \cite{larin}, which preserve
the chiral symmetry in the massless quark limit by the addition of
supplementary counterterms and fulfill the non-renormalization theorem
\cite{adlerbardeen} of the ABJ anomaly \cite{ABJ} at vanishing momentum
transfer. The same result can also be obtained with the scheme of
Ref.~\cite{kreimer} that gives up the cyclicity of the traces involving
Clifford matrices. The next-to-next-to-leading order (NNLO) QCD
corrections have been obtained in the 
limit of heavy top quarks (HTL) \cite{ggannlo}. The QCD corrections are
positive and large in total, increasing the MSSM Higgs production cross
sections at the LHC by up to about 100\%. For the top-loop contributions
alone, the (moderate) NNLO corrections in the heavy-top limit (HTL) can
be used consistently. Electroweak corrections are unknown so far.

The leading terms of the relative QCD corrections in the HTL provide a
reasonable approximation for small $\tgb$ up to pseudoscalar Higgs
masses of $\sim 1$ TeV with a maximal deviation of $\sim 25\%$ for
$\tgb\lsim 5$ at NLO in the intermediate mass range \cite{laenen}. The
genuine SUSY--QCD corrections are only known in the limit of heavy SUSY
particles \cite{gganlosqcd, slavich}.  For large values of $\tgb$ they
can be large and approximated by the $\Delta_b$ terms. This work
improves this incomplete status by calculating the full SUSY--QCD
corrections with full virtual quark-, squark- and gluino-mass
dependence, which will contribute to the virtual corrections as
\begin{equation}
C^A = C_{QCD}^A + C_{SQCD}^A
\label{eq:c_sqcd}
\end{equation}
where $C_{QCD}^A$ is the virtual part of the pure QCD corrections. We
will compare the full results for $C^A_{SQCD}$ with the approximate
calculations in the following sections. For the SUSY--QCD corrections we
implement the stop and sbottom sector at the NLO level, although the
squarks do not contribute at LO, and therefore the definition of a renormalization scheme for their parameters is not required.
However, to be in line with the treatment of scalar Higgs production in a future work, where stops and
sbottoms contribute at LO already, we choose the same framework. The
NLO implementation of the stop and sbottom sectors will be discussed in
Section \ref{sc:squark}.

In the opposite limit, where the pseudoscalar Higgs mass is much larger
than the quark mass, the analytical results of the relative QCD
corrections coincide with the SM expressions at the leading and
subleading logarithmic level for both the scalar and pseudoscalar Higgs
bosons up to NLO where the results for small quark masses are known
\cite{higgsqcd}. This coincidence is due to the restoration of the
chiral symmetry in the massless quark limit. The leading double and
subleading logarithms have been resummed recently \cite{penin}.

\section{Pseudoscalar Higgs Decays} \label{sc:adec}
Although pseudoscalar Higgs decays into gluons and photons do not play a
prominent role as for the SM-like light scalar Higgs particle, they can
still reach sizeable branching ratios for smaller values of $\tgb$ so
that they might be accessible at future $e^+e^-$ colliders.

\subsection{$A\to gg$}
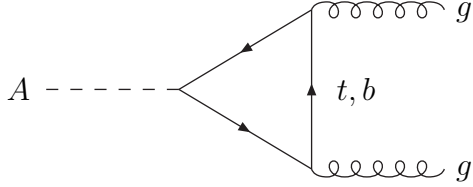
\begin{figure}[hbt]
\begin{center}
\setlength{\unitlength}{1pt}
\begin{picture}(180,100)(0,0)
\DashLine(0,50)(50,50){5}
\ArrowLine(100,20)(100,80)
\ArrowLine(100,80)(50,50)
\ArrowLine(50,50)(100,20)
\Gluon(100,20)(150,20){-3}{5}
\Gluon(100,80)(150,80){3}{5}
\put(-15,46){$A$}
\put(110,46){$t,b$}
\put(155,18){$g$}
\put(155,78){$g$}
\end{picture}  \\
\setlength{\unitlength}{1pt}
\caption{\label{fg:agglodia} \it Typical diagrams contributing to $A\to
gg$ at lowest order.}
\end{center}
\end{figure}

The decay of pseudoscalar Higgs bosons into gluons is loop-induced, see
Fig.~\ref{fg:agglodia}. The dominant contributions originate from top
and bottom loops, while lighter quarks as e.g.~the charm quark yield
contributions at the per-cent or sub-per-cent level only.
The LO expression of the gluonic pseudoscalar Higgs decay reads
\cite{hunter,higgsqcd}
\begin{equation}
\Gamma_{LO} (A\to gg) = \frac{G_F \alpha_s^2 M_A^3}{16\sqrt{2} \pi^3}
\left| \sum_Q g_Q^A A_Q^A (\tau_Q) \right|^2 \;,
\end{equation}
where we adopted the same notation as in Eq.~(\ref{eq:locxn}) using the
same quark form factors as given in Eq.~(\ref{eq:loff}). The NLO QCD and
SUSY--QCD corrections can be cast into the form
\begin{equation}
\Gamma (A\to gg) = \Gamma_{LO} \left\{ 1 + E^A
  \frac{\alpha_s}{\pi}\right\} \;,
\end{equation}
with the NLO coefficient $E^A$ splitting into pure QCD corrections and
genuine SUSY--QCD corrections,
\begin{equation}
E^A = E^A_{QCD} + E^A_{SQCD} \;.
\end{equation}
The QCD part can be expressed as \cite{higgsqcd,bardeen}
\begin{equation}
E^A_{QCD} = \frac{97}{4} - \frac{7}{6} N_F + \Delta_m \;,
\end{equation}
where $\Delta_m$ denotes finite mass effects at NLO \cite{higgsqcd}, and $N_F$ is the number of active light flavors included as final-state quarks as well. For e.g.~$\tgb=1$ the mass effects amount to $\Delta_m \approx 1.3$, if the quark masses are defined as pole masses, but are larger for
increasing values of $\tgb$ due to the rising significance of the bottom
contributions. The expression without $\Delta_m$ corresponds to the
heavy-quark limit of the relative QCD corrections. The coefficient
$E^A_{SQCD}$ coincides with the one for the gluon-fusion cross section
of Eq.~(\ref{eq:c_sqcd}),
\begin{equation}
E^A_{SQCD} = C^A_{SQCD} \;.
\label{eq:a2ggsqcd}
\end{equation}

\subsection{$A\to \gamma\gamma$}
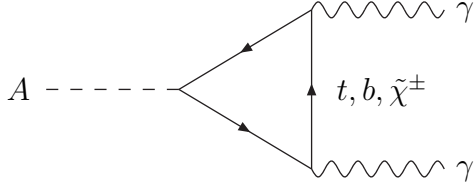
\begin{figure}[hbt]
\begin{center}
\setlength{\unitlength}{1pt}
\begin{picture}(180,100)(0,0)
\DashLine(0,50)(50,50){5}
\ArrowLine(100,20)(100,80)
\ArrowLine(100,80)(50,50)
\ArrowLine(50,50)(100,20)
\Photon(100,20)(150,20){-3}{5}
\Photon(100,80)(150,80){3}{5}
\put(-15,46){$A$}
\put(110,46){$t,b,\tilde \chi^\pm$}
\put(155,18){$\gamma$}
\put(155,78){$\gamma$}
\end{picture}  \\
\setlength{\unitlength}{1pt}
\caption{\label{fg:agagalodia} \it Generic diagrams contributing to $A\to
\gamma\gamma$ at LO.}
\end{center}
\end{figure}

As for the gluonic pseudoscalar Higgs decay, its decay into photon pairs
is a loop-induced process with top and bottom quarks providing
the dominant contributions, but also charginos, see
Fig.~\ref{fg:agagalodia}.
At LO, the pseudoscalar decay width into photon pairs reads
\cite{hunter,higgsqcd}
\begin{equation}
\Gamma_{LO} (A\to \gamma\gamma) = \frac{G_F \alpha^2 M_A^3}{32\sqrt{2} \pi^3}
\left| \sum_f N_{cf} e_f^2 g_f^A A_f^A (\tau_f) + \sum_{\tilde\chi^\pm}
g_{\tilde\chi^\pm}^A A_{\tilde\chi^\pm}^A (\tau_{\tilde\chi^\pm})
\right|^2 \;,
\end{equation}
where $\tilde\chi^\pm$ denotes the two chargino mass eigenstates,
$N_{cf}$ is the color factor of the fermions of charge $e_f$ contributing to the loops.
The LO form factors $A_i^A(\tau_i)~(i=t,b,\tilde\chi^\pm)$ follow the
expressions of Eq.~(\ref{eq:loff}). The chargino-coupling factors are
given by
\begin{equation}
g_{\tilde\chi^\pm_i} = - 2 \frac{M_W}{m_{\tilde\chi^\pm_i}} (S_{ii}
\cos\beta + Q_{ii} \sin\beta) \;,
\end{equation}
with the charge factors $Q_{ii}, S_{ii} (i=1,2)$ given in
Refs.~\cite{haberkane,hunter}. They are related to the mixing angles
between the chargino states $\tilde\chi^\pm_{1,2}$. The NLO QCD and
SUSY--QCD corrections can be defined as a shift of the corresponding LO
quark-form factors,
\begin{equation}
A_Q^A(\tau_Q) \to A_Q^A(\tau_Q) \left\{ 1 + \left[ {\cal D}_{Q,QCD}^A +
{\cal D}_{Q,SQCD}^A \right] \frac{\alpha_s}{\pi} \right\}
\label{eq:d_amp}
\end{equation}
where the pure QCD corrections ${\cal D}_{Q,QCD}$ to the quark form factor
vanish in the heavy-quark limit due to the Adler--Bardeen
\cite{adlerbardeen} theorem for these leading contributions. This means
they are induced by pure quark-mass effects \cite{higgsqcd}. The implementation of the QCD corrections ${\cal D}_{Q,QCD}^A$ follows Ref.~\cite{higgsqcd}, i.e.~the running quark masses
\begin{eqnarray}
\hat m_Q (\mu) & = & \kappa(m_Q) \overline{m}_Q(\mu) \nonumber \\
\kappa(m_Q) & = & 1+\frac{4}{3} \frac{\alpha_{s}(m_Q)}{\pi} + K_Q \left(\frac{\alpha_s(m_Q)}{\pi}\right)^2 + {\cal O}(\alpha_s^3) \nonumber \\
{\overline{m}}_{Q}\,(\mu)&=&{\overline{m}}_{Q}\,(m_{Q})
\,\frac{c\,[\alpha_{s}\,(\mu)/\pi ]}{c\, [\alpha_{s}\,(m_{Q})/\pi ]} \nonumber \\
c(x) & = & \left(\frac{7}{2}\,x\right)^{\frac{4}{7}} \,
[1+1.398x+1.793\,x^{2} - 0.6834\, x^3] \qquad\quad \mbox{for}~m_t < \mu \nonumber \\
& = & \left(\frac{23}{6}\,x\right)^{\frac{12}{23}} \,
[1+1.175x+1.501\,x^{2} + 0.1725\, x^3] \qquad \mbox{for}~m_b < \mu < m_t
\end{eqnarray}
where ${\overline{m}}_{Q}\,(\mu)$ denotes the $\overline{\rm MS}$ mass \cite{mq-msbar} and $K_b = 12.4, K_t = 10.9$ \cite{broadhurst}, are used for the loop-quark masses at the scale $\mu=M_A/2$ such that the relations $M_A = 2 \hat m_Q(m_Q) = 2m_Q~(Q=t,b)$ define the virtual quark thresholds in terms of the quark pole masses $m_Q$. The genuine SUSY--QCD corrections, represented by the coefficient ${\cal D}^A_{Q,SQCD}$, will be discussed in Section \ref{sc:a2gaga}.

\section{Squark Masses and Couplings} \label{sc:squark}
In the following the parametrization of the stop and sbottom sectors
will be described in detail at LO and at NLO starting from the soft
SUSY-breaking parameters, where the 
extension to NLO requires a dedicated scheme choice for our gluon-fusion
calculation. We will follow the set-ups described in
Refs.~\cite{hsqsq,on-shell} with corresponding modifications.

\subsection{Sfermion Masses and Couplings at LO}
Since the scalar sfermion current-eigenstates $\tilde f_{L,R}$, the
super-partners of the the left- and right-handed fermions, mix with each
other, the corresponding mass eigenstates $\tilde f_{1,2}$ are related
to the current eigenstates by a rotation involving the mixing angles
$\theta_f$,
\begin{eqnarray}
\tilde f_1 & = & \tilde f_L \cos\theta_f + \tilde f_R \sin \theta_f \nonumber \\
\tilde f_2 & = & -\tilde f_L\sin\theta_f + \tilde f_R \cos \theta_f \, ,
\label{eq:sfmix}
\end{eqnarray}
These mixing angles grow with the Yukawa couplings of the
corresponding SM fermions, i.e.~mixing effects are in general only
relevant for the third-generation sfermions $\tilde t, \tilde b, \tilde
\tau$. The mass matrix in the current-eigenstate basis is given by
\begin{equation}
{\cal M}_{\tilde f} = \left[ \begin{array}{cc}
\tilde M_{\tilde f_L}^2 + m_f^2 & m_f (A_f-\mu r_f) \\
m_f (A_f-\mu r_f) & \tilde M_{\tilde f_R}^2 + m_f^2 \end{array} \right] \, .
\label{eq:matrix}
\end{equation}
where $r_b = r_\tau = 1/r_t = \tgb$. $A_f$ is the trilinear
sfermion coupling of the soft SUSY-breaking part of the Lagrangian,
while $\mu$ denotes the higgsino mass parameter and $m_f$ the fermion
mass. The parameters $\tilde M_{\tilde f_{L/R}}$ absorb the
corresponding $D$-terms,
\begin{eqnarray}
\tilde M^2_{\tilde f_{L/R}} & = & M^2_{\tilde f_{L/R}} + D_{\tilde f_{L/R}}
\nonumber \\
D_{\tilde f_L} & = & M_Z^2 (I^f_{3L} - e_f \sin^2\theta_W) \cos 2\beta
\nonumber \\
D_{\tilde f_R} & = & M_Z^2 e_f \sin^2\theta_W \cos 2\beta \, ,
\label{eq:dterms}
\end{eqnarray}
with $e_f$ being the electric charge of the sfermion, and $I_{3L}$ its
third isospin component, $\theta_W$ denotes the Weinberg angle and $M_{\tilde f_{L/R}}$ are the sfermion mass parameters of the soft
SUSY-breaking part of the Lagrangian. Hence, the mixing angles are
determined from
\begin{equation}
\sin 2\theta_f = \frac{2m_f (A_f-\mu r_f)}{m_{\tilde f_1}^2 - m_{\tilde f_2}^2}
~~~,~~~
\cos 2\theta_f = \frac{\tilde M_{\tilde f_L}^2 - \tilde M_{\tilde f_R}^2}
{m_{\tilde f_1}^2 - m_{\tilde f_2}^2}
\label{eq:sqmix0}
\end{equation}
and the squark-eigenstate masses acquire the form
\begin{equation}
m_{\tilde f_{1,2}}^2 = m_f^2 + \frac{1}{2}\left[ \tilde M_{\tilde f_L}^2 +
\tilde M_{\tilde f_R}^2 \mp \sqrt{(\tilde M_{\tilde f_L}^2 -
\tilde M_{\tilde f_R}^2)^2 + 4m_f^2 (A_f - \mu r_f)^2} \right] \, .
\label{eq:losqmass}
\end{equation}
In the current-eigenstate basis, the neutral Higgs couplings to
sfermions are given by
\begin{eqnarray}
g_{\tilde f_L \tilde f_L}^\Phi & = & m_f^2 g_1^\Phi + M_Z^2 (I_{3f}
- e_f\sin^2\theta_W) g_2^\Phi \nonumber \\
g_{\tilde f_R \tilde f_R}^\Phi & = & m_f^2 g_1^\Phi + M_Z^2 e_f\sin^2\theta_W
g_2^\Phi \nonumber \\
g_{\tilde f_L \tilde f_R}^\Phi & = & \frac{m_f}{2} (\mu g_3^\Phi
- A_f g_4^\Phi) \, ,
\label{eq:hsfcouprl}
\end{eqnarray}
where the couplings $g_i^\Phi~(i=1,\ldots,4)$ are specified in Table
\ref{tb:hsfcoup}. In case of the scalar Higgs bosons $h,H$ the couplings
to sfermions are symmetric, i.e.  $g_{\tilde f_R \tilde f_L}^{h,H} =
g_{\tilde f_L \tilde f_R}^{h,H}$, while for the pseudoscalar Higgs boson
$A$ the diagonal couplings $g_{\tilde f_L \tilde f_L}^A$ and $g_{\tilde
f_R \tilde f_R}^A$ vanish and the off-diagonal couplings are
antisymmetric, $g_{\tilde f_R \tilde f_L}^A=-g_{\tilde f_L \tilde
f_R}^A$. The physical Higgs couplings to the sfermion mass eigenstates
$\tilde f_{1,2}$ read
\begin{eqnarray}
g_{\tilde f_1 \tilde f_1}^{h,H} & = & g_{\tilde f_L \tilde f_L}^{h,H}
\cos^2\theta_f + g_{\tilde f_R \tilde f_R}^{h,H} \sin^2\theta_f +
g_{\tilde f_L \tilde f_R}^{h,H} \sin 2\theta_f \nonumber \\
g_{\tilde f_2 \tilde f_2}^{h,H} & = & g_{\tilde f_L \tilde f_L}^{h,H}
\sin^2\theta_f + g_{\tilde f_R \tilde f_R}^{h,H} \cos^2\theta_f -
g_{\tilde f_L \tilde f_R}^{h,H} \sin 2\theta_f \nonumber \\
g_{\tilde f_1 \tilde f_2}^{h,H} & = & g_{\tilde f_2 \tilde f_1}^{h,H}
= \frac{1}{2}(g_{\tilde f_R \tilde f_R}^{h,H}
- g_{\tilde f_L \tilde f_L}^{h,H}) \sin 2\theta_f + g_{\tilde f_L \tilde
  f_R}^{h,H} \cos 2\theta_f \nonumber \\
g_{\tilde f_1 \tilde f_1}^A & = & g_{\tilde f_2 \tilde f_2}^A = 0
\nonumber \\
g_{\tilde f_1 \tilde f_2}^A & = & -g_{\tilde f_2 \tilde f_1}^A =
g_{\tilde f_L \tilde f_R}^A \, .
\label{eq:couprot}
\end{eqnarray}
Next, we will discuss the extension of the stop and sbottom sectors to
the NLO SUSY--QCD level.
\begin{table}[hbt]
\renewcommand{\arraystretch}{1.5}
\begin{center}
\begin{tabular}{|l|c||c|c|c|c|} \hline
$\tilde f$ & $\Phi$ & $g^\Phi_1$ & $g^\Phi_2$ & $g^\Phi_3$ & $g^\Phi_4$ \\
\hline \hline
& $h$ & $\cos\alpha/\sin\beta$ & $-\sin(\alpha+\beta)$ &
$-\sin\alpha/\sin\beta$ & $\cos\alpha/\sin\beta$ \\
$\tilde u$ & $H$ & $\sin\alpha/\sin\beta$ & $\cos(\alpha+\beta)$ &
$\cos\alpha/\sin\beta$ & $\sin\alpha/\sin\beta$ \\
& $A$ & 0 & 0 & 1 & $-1/\tgb$ \\ \hline
& $h$ & $-\sin\alpha/\cos\beta$ & $-\sin(\alpha+\beta)$ &
$\cos\alpha/\cos\beta$ & $-\sin\alpha/\cos\beta$ \\
$\tilde d$ & $H$ & $\cos\alpha/\cos\beta$ & $\cos(\alpha+\beta)$ &
$\sin\alpha/\cos\beta$ & $\cos\alpha/\cos\beta$ \\
& $A$ & 0 & 0 & 1 & $-\tgb$ \\ \hline
\end{tabular} 
\renewcommand{\arraystretch}{1.2}
\caption[]{\label{tb:hsfcoup}
\it Coefficients of the neutral MSSM Higgs couplings to sfermion pairs.
The symbols $\tilde u,\tilde d$ denote up- and down-type sfermions.}
\end{center}
\end{table}

\subsection{Stops and Sbottoms at NLO}
At NLO, we will introduce the soft SUSY-breaking parameters in the
$\overline{\rm MS}$ scheme, i.e.~we will start from the soft
supersymmetry-breaking parameters $\overline{M}_{\LF{\tilde Q_{L,R}}}(Q_0)$
and $\overline{A}_\LF{Q}(Q_0)$ at the input scale $Q_0$ which will in general be
the SUSY scale, i.e.~the average size of the left- and right-handed soft
SUSY-breaking mass parameters. The benchmark scenarios of
Ref.~\cite{bench}, however, are defined in the on-shell scheme of all
involved input parameters. Thus, we will describe how we are
implementing the relation between the $\overline{\rm MS}$ and the on-shell
parameters.

The bottom and top masses involved in the sbottom and stop mass matrices
have to be chosen such that large higher-order corrections to their
entries are avoided. We have chosen the top pole mass and a derived bottom mass for
the sbottom mass matrix according to Refs.~\cite{on-shell}.  At LO, the
stop/sbottom mass matrix is then given by ($q=t,b$)
\begin{equation}
{\cal M}_{\LF{\tilde Q}} = \left[ \begin{array}{cc}
\tilde{\overline{M}}_{\LF{\tilde Q}_L}^2(Q_0) + m_\LF{Q}^2
& m_\LF{Q} [\bar{A}_\LF{Q}(Q_0)-\mu r_\LF{Q}] \\
m_\LF{Q} [\bar{A}_\LF{Q}(Q_0)-\mu r_\LF{Q}]
& \tilde{\overline{M}}_{\LF{\tilde Q}_R}^2(Q_0) + m_\LF{Q}^2
\end{array} \right] \, ,
\end{equation}
where $m_t$ is the top pole mass and $m_b$ is the derived bottom mass as
will be discussed in the following. The $D$-terms $D_{\tilde Q_{L/R}}$
have again been absorbed in the soft SUSY-breaking parameters,
$\tilde{\overline{M}}_{\tilde Q_{L/R}}(Q_0)$,
\begin{equation}
\tilde{\overline{M}}^2_{\tilde Q_{L/R}}(Q_0) = \overline{M}^2_{\tilde
Q_{L/R}}(Q_0) + D_{\tilde Q_{L/R}} \, .
\label{eq:dterms_b}
\end{equation}
The diagonal and off-diagonal entries of the stop/sbottom mass matrix
are corrected at higher orders. We absorb the radiative corrections to
the diagonal matrix elements in shifted soft mass parameters, $M_{\tilde
Q_{L/R}}$,
\begin{equation}
M_{\tilde Q_{L/R}}^2 = \overline{M}^2_{\tilde Q_{L/R}}(Q_0)
+ \Delta \overline{M}_{\tilde Q_{L/R}}^2 \qquad , \qquad 
\tilde M_{\tilde Q_{L/R}}^2 = \tilde{\overline{M}}^2_{\tilde Q_{L/R}}(Q_0)
+ \Delta \overline{M}_{\tilde Q_{L/R}}^2 \, ,
\label{eq:shift}
\end{equation}
while the corrections to the off-diagonal entries will be absorbed in
shifted soft trilinear couplings,
\begin{equation}
A_Q = \overline{A}_Q(Q_0) + \Delta \overline{A}_Q \, .
\label{eq:a-shift}
\end{equation}
The shifted parameters are related to the radiative corrections to the
mixing angles and stop/sbottom masses in order to arrive at simple
tree-level like expressions at NLO for the stop/sbottom parameters. On
the other hand, these shifted parameters correspond to the on-shell
scheme introduced in Refs.~\cite{on-shell} and thus have to coincide with
the input values of the chosen benchmark scenario.

\subsubsection{Stops}
Starting from the on-shell parameters the treatment of the stop sector
is identical to the LO level discussed before. The relation of the
on-shell to the $\overline{\rm MS}$ parameters, however, is affected by the
NLO corrections.

At tree-level, the mixing angle $\tilde \theta_Q$ is derived from
\begin{equation}
\sin 2\tilde\theta_Q = \frac{2 m_Q [\overline{A}_Q(Q_0)-\mu r_Q]}{ m_{\tilde
Q_1}^2 - m_{\tilde Q_2}^2} \quad , \quad
\cos 2\tilde\theta_Q = \frac{\tilde{\overline{M}}_{\tilde Q_L}^2(Q_0)
- \tilde{\overline{M}}_{\tilde Q_R}^2(Q_0)}
{m_{\tilde Q_1}^2 - m_{\tilde Q_2}^2} \, ,
\label{eq:sqmix}
\end{equation}
where the tree-level squark masses $m_{\tilde q_{1/2}}$ according to
Eq.~(\ref{eq:losqmass}) have been used\footnote{The standard range for the squark mixing angle is chosen between 0 and $\pi$.}.

The masses of the stop/sbottom
mass eigenstates acquire radiative corrections,
\begin{eqnarray}
m_{\tilde Q_{1/2}}^2 & = & m_Q^2 + \frac{1}{2}\left[
\tilde{\overline{M}}_{\tilde Q_L}^2(Q_0) + \tilde{\overline{M}}_{\tilde
Q_R}^2(Q_0) \right. \nonumber \\
& & \left. \hspace*{1.5cm} \mp \sqrt{[\tilde{\overline{M}}_{\tilde
Q_L}^2(Q_0) - \tilde{\overline{M}}_{\tilde Q_R}^2(Q_0)]^2 + 4
m_Q^2 [\overline{A}_Q(Q_0) - \mu r_Q]^2} \right] + \Delta m_{\tilde
Q_{1/2}}^2 \nonumber \\
\Delta m_{\tilde Q_{1/2}}^2 & = & \Sigma_{11/22}(m_{\tilde Q_{1/2}}^2) +
\delta \hat m_{\tilde Q_{1/2}}^2 \, .
\label{eq:sqmass}
\end{eqnarray}
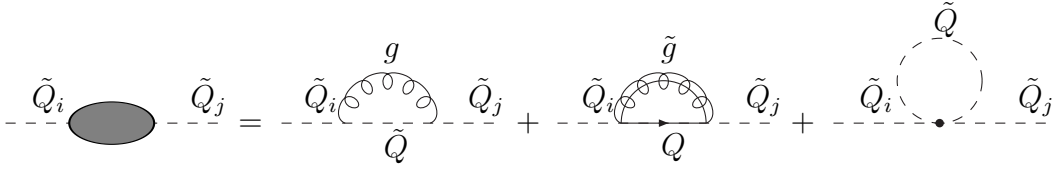
\begin{figure}[t!]
\begin{center}
\SetScale{0.8}
\begin{picture}(130,40)(130,-10)
\DashLine(0,0)(100,0){5}
\GOval(50,0)(10,20)(0){0.5}
\DashLine(130,0)(230,0){5}
\GlueArc(180,0)(20,0,180){4}{5}
\DashLine(260,0)(360,0){5}
\ArrowLine(290,0)(330,0)
\CArc(310,0)(20,0,180)
\GlueArc(310,0)(20,0,180){4}{5}
\DashLine(390,0)(490,0){5}
\DashCArc(440,20)(20,0,360){5}
\Vertex(440,0){2}
\put(88,-3){$=$}
\put(193,-3){$+$}
\put(297,-3){$+$}
\put(10,6){$\tilde Q_i$}
\put(70,6){$\tilde Q_j$}
\put(113,6){$\tilde Q_i$}
\put(143,-13){$\tilde Q$}
\put(143,25){$g$}
\put(175,6){$\tilde Q_j$}
\put(218,6){$\tilde Q_i$}
\put(248,-12){$Q$}
\put(248,25){$\tilde g$}
\put(280,6){$\tilde Q_j$}
\put(323,6){$\tilde Q_i$}
\put(351,35){$\tilde Q$}
\put(383,6){$\tilde Q_j$}
\end{picture} \\
\caption{\label{fg:sqself} \it One-loop contributions to the squark
self-energies.}
\vspace*{-0.5cm}
\end{center}
\end{figure}
The self-energies $\Sigma_{11/22}$ of the stops/sbottoms can be derived 
from the diagrams in Fig.~\ref{fg:sqself}, 
\begin{eqnarray}
\Sigma_{11/22}(m_{\tilde Q_{1/2}}^2) & = & C_F\frac{\alpha_s}{\pi}
\frac{1}{4} \left\{
-(1+\cos^2 2\tilde\theta_Q) A_0(m_{\tilde Q_{1/2}}) - \sin^2 2\tilde\theta_Q
A_0(m_{\tilde Q_{2/1}}) \right.  \nonumber \\
& & + 2A_0(M_\sgl) + 2A_0(m_Q) + 4 m_{\tilde Q_{1/2}}^2
B_0(m_{\tilde Q_{1/2}}^2;0,m_{\tilde Q_{1/2}}) \nonumber \\
& + & \left. \!\!\!\!\! 2\!\left[ M_\sgl^2 + m^2_Q - m_{\tilde
Q_{1/2}}^2 \!\! \mp 2M_\sgl m_Q \sin 2\tilde\theta_Q \right] \!
B_0(m_{\tilde Q_{1/2}}^2;M_\sgl, m_Q) \right\} \, ,
\label{eq:sigii}
\end{eqnarray}
where $M_{\tilde g}$ denotes the gluino mass and the scalar one-loop integrals are defined as
($n=4-2\epsilon$) \cite{passvelt} 
\begin{eqnarray}
A_0(m) & = & \int \frac{d^n k}{(2\pi)^n} \frac{-i(4\pi)^2
\bar\mu^{2\epsilon}}{k^2-m^2} \nonumber \\
B_0(p^2;m_1,m_2) & = & \int \frac{d^n k}{(2\pi)^n} \frac{-i(4\pi)^2
\bar\mu^{2\epsilon}}{[k^2-m_1^2][(k+p)^2-m_2^2]} \nonumber \\
B_1(p^2;m_1,m_2) & = & \frac{1}{2p^2}\left\{ A_0(m_1)-A_0(m_2) -
(p^2+m_1^2-m_2^2) B_0(p^2;m_1,m_2) \right\} \, .
\end{eqnarray}
The scale $\bar \mu$ denotes the 't Hooft mass of dimensional
regularization. The mass counterterms $\delta \hat m_{\tilde Q_{1,2}}^2$ of
Eq.~(\ref{eq:sqmass}) are related to the counterterms of the input
parameters,
\begin{eqnarray}
\delta \hat m_{\tilde Q_{1/2}}^2 & = & 2 m_Q\delta m_Q +
\frac{1}{2}\left\{ \delta \overline{M}_{\tilde Q_L}^2 + \delta
\overline{M}_{\tilde Q_R}^2 \pm \left[ (\delta \overline{M}_{\tilde
Q_L}^2-\delta \overline{M}_{\tilde Q_R}^2) \cos 2\tilde \theta_Q \right.
\right. \nonumber \\
& & \left. \left. \hspace*{1.5cm} + \left( \frac{\delta m_Q}{m_Q}
+ \frac{\delta \overline{A}_Q}{\overline{A}_Q(Q_0)-\mu r_Q} \right)
(m_{\tilde Q_1}^2 - m_{\tilde Q_2}^2) \sin^2 2\tilde \theta_Q \right]
\right\} \nonumber \\
& = & - C_F\frac{\alpha_s}{\pi} \Gamma(1+\epsilon) (4\pi)^\epsilon
\left\{ \frac{1}{\epsilon} + \log\frac{\bar \mu^2}{Q_0^2} \right\}
\left\{M^2_\sgl \mp M_\sgl m_Q \sin 2\tilde \theta_Q) \right\}
\nonumber \\
& & + \frac{\delta m_Q}{m_Q} \left\{ 2 m^2_Q \mp
\frac{1}{2}(m_{\tilde Q_2}^2 - m_{\tilde Q_1}^2) \sin^2 2\tilde
\theta_Q \right\} \, ,
\label{eq:dmsq}
\end{eqnarray}
using the tree-level mixing angle $\tilde \theta_Q$ of
Eq.~(\ref{eq:sqmix}), and $C_F=4/3$. The counterterms of the parameters
$\overline{M}_{\sq_{L/R}}^2(Q_0)$ and $\overline{A}_Q (Q_0)$ are defined
in the $\overline{\rm MS}$ scheme,
\begin{eqnarray}
\delta \overline{M}^2_{\sq_{L/R}} & = & -C_F \frac{\alpha_s}{\pi}
\Gamma(1+\epsilon) (4\pi)^\epsilon M^2_\sgl \left\{ \frac{1}{\epsilon} +
\log \frac{\bar \mu^2}{Q_0^2} \right\} \nonumber \\
\delta \overline{A}_Q & = & C_F \frac{\alpha_s}{\pi} \Gamma(1+\epsilon)
(4\pi)^\epsilon M_\sgl \left\{ \frac{1}{\epsilon} + \log \frac{\bar
\mu^2}{Q_0^2} \right\} \, .
\label{eq:daq}
\end{eqnarray}
The counterterm of the pole quark mass $m_q$ is given by
\begin{eqnarray}
\frac{\delta m_Q}{m_Q} & = & -C_F\frac{\alpha_s}{4\pi}
\left\{ \frac{A_0(m_Q)}{m_Q^2} + 2 B_0(m_Q^2;0,m_Q) - 1
+B_1(m_Q^2;M_\sgl,m_{\tilde Q_1}) + B_1(m_Q^2;M_\sgl,m_{\tilde
Q_2}) \right.  \nonumber \\
& & \left. \hspace*{0.7cm} + \delta_{SUSY}
+2 M_\sgl (A_Q-\mu r_Q) \frac{B_0(
m_Q^2;M_\sgl,m_{\tilde Q_1}) - B_0(m_Q^2;M_\sgl,m_{\tilde
Q_2})}{m^2_{\tilde Q_1}-m^2_{\tilde Q_2}}\right\} \, ,
\label{eq:dmq}
\end{eqnarray}
where $\delta_{SUSY}=1/3$ is a finite counterterm required to restore
the supersymmetric relation between the Higgs-boson couplings to quarks
and squarks within dimensional regularization \cite{susyrest}. 
The definition of the mixing angle $\tilde\theta_Q$ in Eq.~(\ref{eq:sqmix})
corresponds to the following counterterm at NLO,
\begin{eqnarray}
\delta\tilde\theta_Q & = & \frac{{\rm tg}~2\tilde\theta_Q}{2} \left\{
\frac{\delta m_Q}{m_Q} + \frac{\delta \overline{A}_Q}
{\overline{A}_Q(Q_0)-\mu r_Q}
 - \frac{\delta m_{\tilde Q_1}^2 - \delta m_{\tilde
Q_2}^2}{m_{\tilde Q_1}^2 - m_{\tilde Q_2}^2} \right\} \, ,
\nonumber \\
\delta m_{\tilde Q_{1/2}}^2 & = & -\Sigma_{11/22}(m_{\tilde Q_{1/2}}^2)
\, .
\label{eq:sqmassct}
\end{eqnarray}
However, this mixing angle definition induces artificial singularities
in physical observables for stop/sbottom masses $m_{\tilde q_{1,2}}$
close to each other \cite{h2sqcorr}. To avoid such singularities, the
mixing angle of the squark fields has been renormalized via the
anti-Hermitian (on-shell) counterterm \cite{h2sqcorr},
\begin{equation}
\delta\theta_Q = -\frac{1}{2} \frac{\Re \Sigma_{12}(m_{\tilde
Q_1}^2) - \Re \Sigma_{12}(m_{\tilde Q_2}^2)}{m_{\tilde Q_1}^2
- m_{\tilde Q_2}^2} \;,
\label{eq:dtheta}
\end{equation}
with the off-diagonal part $\Sigma_{12}$ of the stop/sbottom
self-energy (see Fig.~\ref{fg:sqself}) describing transitions from the
first to the second mass eigenstate or {\it vice versa},
\begin{equation}
\Sigma_{12}(m^2) = - C_F\frac{\alpha_s}{\pi} \left\{ M_\sgl m_Q
B_0(m^2;M_\sgl,m_Q) + \frac{\sin 2\tilde\theta_Q}{4} \Big[
A_0(m_{\tilde Q_2}) - A_0(m_{\tilde Q_1}) \Big] \right\} \cos
2\tilde\theta_Q \, . \!\!\!
\label{eq:sig12}
\end{equation}
For the mixing angle $\tilde\theta_Q$ of Eq.~(\ref{eq:sqmix}),
this implies a finite shift $\Delta \tilde\theta_Q$,
\begin{equation}
\theta_Q = \tilde\theta_Q + \Delta \tilde\theta_Q \qquad , \qquad
\Delta \tilde\theta_Q = \delta \tilde\theta_Q - \delta \theta_Q
\label{eq:sqmix1}
\end{equation}
that will be absorbed in the shifted $A_Q$ value of
Eq.~(\ref{eq:a-shift}). This shift defines the relation between the
on-shell coupling $A_Q$ and the $\overline{\rm MS}$ one
$\overline{A}_Q(Q_0)$.

Using the NLO corrected squark pole masses of Eq.~(\ref{eq:sqmass}) and
the radiatively corrected mixing angle $\theta_q$, the shifted
(on-shell) squared soft SUSY-breaking squark mass parameters $\tilde
M^2_{\tilde Q_{L/R}} = \tilde{\overline{M}}^2_{\tilde Q_{L/R}}(Q_0) +
\Delta \overline{M}^2_{\tilde Q_{L/R}}$ can be obtained from the sum
rules,
\begin{eqnarray}
\tilde M^2_{\tilde Q_L} & = & M_{\tilde Q_L}^2 + D_{\tilde Q_L} =
m^2_{\tilde Q_1} \cos^2 \theta_Q + m^2_{\tilde Q_2} \sin^2
\theta_Q - m_Q^2 \nonumber \\
\tilde M^2_{\tilde Q_R} & = & M_{\tilde Q_R}^2 + D_{\tilde Q_R} =
m^2_{\tilde Q_1} \sin^2 \theta_Q + m^2_{\tilde Q_2} \cos^2
\theta_Q - m_Q^2 
\label{eq:mlr}
\end{eqnarray}
while the shifted (on-shell) trilinear couplings $A_Q$ are derived from
the relation
\begin{eqnarray}
A_Q = \frac{m_{\tilde Q_1}^2 - m_{\tilde Q_2}^2}{2m_Q} \sin 2\theta_Q +
\mu r_Q \, .
\label{eq:a-par}
\end{eqnarray}
In terms of these shifted (on-shell) parameters the radiatively
corrected squark masses and mixing angles are given by LO-like
expressions,
\begin{eqnarray}
m_{\tilde Q_{1/2}}^2 & = & m_Q^2 + \frac{1}{2}\left[ \tilde{M}_{\tilde
Q_L}^2 + \tilde{M}_{\tilde Q_R}^2 \mp \sqrt{(\tilde{M}_{\tilde Q_L}^2 -
\tilde{M}_{\tilde Q_R}^2)^2 + 4 m_Q^2 (A_Q - \mu r_Q)^2} \right] \nonumber \\
\sin 2\theta_Q & = & \frac{2 m_Q (A_Q-\mu r_Q)}{ m_{\tilde Q_1}^2
- m_{\tilde Q_2}^2} \quad , \quad \cos 2\theta_Q = \frac{\tilde
M_{\tilde Q_L}^2 - \tilde M_{\tilde Q_R}^2} {m_{\tilde Q_1}^2 -
m_{\tilde Q_2}^2} \, .
\label{eq:shiftpar}
\end{eqnarray}
The scale of the strong coupling constants $\alpha_s$ in
Eqs.~(\ref{eq:sigii}, \ref{eq:dmsq}, \ref{eq:daq}, \ref{eq:dmq},
\ref{eq:sig12}) has been identified with the input scale $Q_0$.

These relations have been used for the determination of the
$\overline{\rm MS}$ parameters $\tilde{\overline{M}}_{\tilde
Q_{L/R}}^2(Q_0)$ and $\overline{A}_q(Q_0)$ iteratively until the on-shell
parameters agreed with the input value of the chosen benchmark scenario.

\subsubsection{Sbottoms}
The procedure described for the stops is necessary to obtain the
$\overline{\rm MS}$ parameter $\overline{M}_{\tilde t_{L}}(Q_0)$
that by virtue of the SU(2) gauge symmetry is identified with the
$\overline{\rm MS}$ parameter $\overline{M}_{\tilde b_L}(Q_0)$,
\begin{equation}
\overline{M}_{\tilde t_{L}}(Q_0) =
\overline{M}_{\tilde b_{L}}(Q_0) \;.
\end{equation}
Due to potentially large $\tgb$-enhanced contributions in the sbottom
sector the procedure has to be modified. This modification addresses the
treatment and renormalization of the bottom mass $m_b$ and of the trilinear
coupling $A_b$. Therefore, the bottom mass is not introduced as the pole mass, but
as a derived quantity, since it represents the contribution of the
bottom Yukawa coupling to the sbottom sector. To achieve a working
scheme, we are starting from Eq.~(\ref{eq:shiftpar}) for the mixing angle
that at NLO is still defined via the anti-Hermitian counterterm of
Eq.~(\ref{eq:dtheta}). The trilinear coupling $A_b$, however, is now
defined from the proper $A\tilde b_1 \tilde b_2$ vertex \cite{on-shell}.
This definition avoids large $\tgb$-enhanced contributions in the
renormalization of $A_b$. The bottom mass $m_b$ entering the sbottom
mixing matrix is then treated as a derived quantity. This leads to the
explicit counterterms,
\begin{eqnarray}
\label{eq:mbct}
\delta A_b & = & -\frac{s_\beta c_\beta}{\mu} (A_b-\mu\tgb) \left( A_b +
\frac{\mu}{\tgb}\right) \left\{ F - 2\frac{c_{2\theta_b}}{s_{2\theta_b}}
\delta\theta_b - \frac{\delta m_{\tilde b_1}^2 - \delta m_{\tilde
b_2}^2}{m_{\tilde b_1}^2 - m_{\tilde b_2}^2} \right\} \\
\frac{\delta \hat m_b}{m_b} & = & \left\{ 1+\frac{s_\beta c_\beta}{\mu}
(A_b-\mu\tgb) \right\} F - \frac{s_\beta c_\beta}{\mu} (A_b-\mu\tgb)
\left\{ 2\frac{c_{2\theta_b}}{s_{2\theta_b}}
\delta\theta_b + \frac{\delta m_{\tilde b_1}^2 - \delta m_{\tilde
b_2}^2}{m_{\tilde b_1}^2 - m_{\tilde b_2}^2} \right\} \;, \nonumber
\end{eqnarray}
where the term $F$ is defined as \cite{sushi}
\begin{eqnarray}
F & = & f(m_{\tilde b_1}^2,m_{\tilde b_2}^2) + f(m_{\tilde
b_2}^2,m_{\tilde b_1}^2) \nonumber \\
f(m_1^2,m_2^2) & = & -\frac{C_F}{2} \frac{\alpha_s}{\pi} \left\{
-\frac{M_{\tilde g}}{A_b+\mu\cot\beta} B_0(m_1^2;M_{\tilde g}, m_b)
\right. \nonumber \\
& & \left. + \frac{m_1^2}{m_1^2-m_2^2} \left[
2B_0(m_1^2;0,m_1) - \frac{m_1^2-M_{\tilde g}^2-m_b^2}{m_1^2}
B_0(m_1^2;M_{\tilde g}, m_b) \right] \right\}.
\end{eqnarray}
The derived bottom mass $\hat m_b$ is then determined as
\begin{equation}
\hat m_b = \overline{m}_b (Q_0) -\delta \hat m_b + \delta
\overline{m}_b \;, \label{eq:derivedbotmass}
\end{equation}
where $\overline{m}_b (Q_0)$ denotes the $\overline{\rm MS}$ bottom mass at
the input scale $Q_0$, $\delta \hat m_b$ the counterterm of
Eq.~(\ref{eq:mbct}) and $\delta \overline{m}_b$ the $\overline{\rm MS}$
counterterm of the bottom mass,
\begin{eqnarray}
\frac{\delta \overline{m}_b}{m_b} & = & -C_F\frac{\alpha_s}{\pi}
\Gamma(1+\epsilon) (4\pi)^\epsilon \frac{3}{4} \left\{\frac{1}{\epsilon}
+\log \frac{\bar \mu^2}{Q_0^2} + \delta_{SUSY} \right\}
\nonumber \\
& - & C_F \frac{\alpha_s}{4\pi} \left\{ B_1[
m_b^2;M_\sgl,m_{\tilde b_1}] + B_1[m_b^2;M_\sgl,m_{\tilde
b_2}] \right. \nonumber \\
& & \left. \hspace*{0.7cm} +2 M_\sgl (A_b-\mu\tgb) \frac{B_0[
m_b^2;M_\sgl,m_{\tilde b_1}] - B_0[m_b^2;M_\sgl,m_{\tilde
b_2}]}{m^2_{\tilde b_1}-m^2_{\tilde b_2}}\right\} \, ,
\end{eqnarray}
where $\delta_{SUSY}=1/3$ is a SUSY-restoring counterterm. This
$\overline{\rm MS}$ counterterm defines the running bottom mass with decoupled
SUSY contributions, i.e.~the running bottom mass of the SM. The derived
bottom mass $\hat m_b$ is then used for the sbottom mixing matrix
throughout. In the analogous way we determine the $\overline{\rm MS}$ value
$\overline{A}_b(Q_0)$ of the trilinear coupling, but this will not be
used in our analysis.

The shifted (on-shell) sbottom mass parameters $\tilde M_{\tilde
b_{L/R}}$ are finally determined from the corresponding sum rules of
Eq.~(\ref{eq:mlr}). This
set-up of the sbottom sector is then used for 
iteration until the on-shell parameter $\tilde M_{\tilde b_{R}}$ agrees
with the input parameter of the benchmark scenario.

An alternative approach is provided by a purely fixed-order
implementation of the difference between $M_{\tilde b_{L}}$ and
$M_{\tilde t_{L}}$,
\begin{eqnarray}
M^2_{\tilde b_{L}} & = & M^2_{\tilde t_{L}} + \Delta M^2_L
\nonumber \\
\Delta M_L^2 & = & \delta M^2_{\tilde t_{L}} - \delta
M^2_{\tilde b_{L}} \nonumber \\
\delta M^2_{\tilde q_{L}} & = & c^2_{\theta_q} \delta
m^2_{\tilde q_{1}} + s^2_{\theta_q} \delta
m^2_{\tilde q_{2}} - (m^2_{\tilde q_{1}} - m^2_{\tilde q_{2}})
s_{2\theta_q} \delta\theta_q - 2m_q\delta m_q \;,
\end{eqnarray}
with $q=t,b$. The counterterms $\delta m_{\tilde q_{1/2}}$ are given in
Eq.~(\ref{eq:sqmassct}), the counterterm $\delta\theta_q$ in
Eq.~(\ref{eq:dtheta}) and the counterterm $\delta m_q$ in
Eq.~(\ref{eq:dmq}) for the top pole mass $m_q=m_t$ and in
Eq.~(\ref{eq:mbct}) for the (derived) bottom mass $m_q=\hat m_b$. This
approach does not require any iteration, since the on-shell parameters
of the benchmark scenario can immediately be used to derive the
parameters of the sbottom sector. We have compared both approaches and
found agreement of the sbottom parameters at the few-per-mille
level.

\subsubsection{Higgs Couplings to Stops and Sbottoms}
The NLO neutral Higgs couplings to squarks in the current-eigenstate
basis are given by
\begin{eqnarray}
g_{\tilde Q_L \tilde Q_L}^\Phi & = & m_Q^2 g_1^\Phi +
M_Z^2 (I_{3Q} - e_Q\sin^2\theta_W) g_2^\Phi \nonumber \\
g_{\tilde Q_R \tilde Q_R}^\Phi & = & m_Q^2 g_1^\Phi +
M_Z^2 e_Q\sin^2\theta_W g_2^\Phi \nonumber \\
g_{\tilde Q_L \tilde Q_R}^\Phi & = & \frac{m_Q}{2}
\left[ \mu g_3^\Phi - A_Q g_4^\Phi \right] \;,
\label{eq:hsbsbcoup}
\end{eqnarray}
with the on-shell trilinear couplings $A_Q$ and the couplings $g_i^\Phi$
of Table \ref{tb:hsfcoup}. The quark mass $m_Q$ denotes either the top
pole mass in the stop case or the derived bottom mass $\hat m_b$ for the
sbottom sector. The related couplings to the stop/sbottom mass
eigenstates $\tilde Q_{1,2}$ are derived by the rotations according to
Eq.~(\ref{eq:couprot}) by the radiatively corrected mixing angle
$\theta_Q$. For pseudoscalar Higgs bosons, we obtain vanishing diagonal
couplings $g_{\tilde Q_L \tilde Q_L}^A = g_{\tilde Q_R \tilde Q_R}^A =
0$ and non-vanishing off-diagonal couplings $g_{\tilde Q_1 \tilde Q_2}^A
= -g_{\tilde Q_2 \tilde Q_1}^A = g_{\tilde Q_L \tilde Q_R}^A$ at the NLO
level as at LO.

\section{SUSY--QCD corrections at NLO} \label{sc:nlo}
\begin{figure}[htb]
\begin{center}
\begin{picture}(130,90)(15,0)
\Gluon(10,20)(50,20){-3}{4}
\Gluon(10,80)(50,80){3}{4}
\ArrowLine(50,20)(50,80)
\ArrowLine(50,80)(70,65)
\ArrowLine(70,35)(50,20)
\DashLine(70,65)(90,50){5}
\DashLine(90,50)(70,35){5}
\Line(70,65)(70,35)
\Gluon(70,65)(70,35){-3}{3}
\DashLine(90,50)(130,50){5}
\put(0,18){$g$}
\put(0,78){$g$}
\put(35,46){$Q$}
\put(60,46){$\tilde g$}
\put(80,60){$\tilde Q$}
\put(120,36){$A$}
\end{picture}
\begin{picture}(130,90)(0,0)
\Gluon(10,20)(50,20){-3}{4}
\Gluon(10,80)(50,80){3}{4}
\DashLine(50,50)(50,80){5}
\DashLine(50,80)(70,65){5}
\ArrowLine(70,65)(90,50)
\ArrowLine(90,50)(50,20)
\ArrowLine(50,20)(50,50)
\Line(70,65)(50,50)
\Gluon(70,65)(50,50){-3}{3}
\DashLine(90,50)(130,50){5}
\put(0,18){$g$}
\put(0,78){$g$}
\put(35,60){$\tilde Q$}
\put(60,46){$\tilde g$}
\put(80,60){$Q$}
\put(120,36){$A$}
\end{picture}
\begin{picture}(130,90)(-15,0)
\Gluon(10,20)(50,20){-3}{4}
\Gluon(10,80)(50,80){3}{4}
\ArrowLine(50,65)(50,80)
\ArrowLine(50,20)(50,35)
\ArrowLine(50,80)(90,50)
\ArrowLine(90,50)(50,20)
\GlueArc(50,50)(15,90,270){3}{5}
\CArc(50,50)(15,90,270)
\DashCArc(50,50)(15,270,450){5}
\DashLine(90,50)(130,50){5}
\put(0,18){$g$}
\put(0,78){$g$}
\put(67,46){$\tilde Q$}
\put(20,46){$\tilde g$}
\put(80,60){$Q$}
\put(120,36){$A$}
\end{picture} \\
\begin{picture}(130,90)(15,0)
\Gluon(10,20)(50,20){-3}{4}
\Gluon(10,80)(50,80){3}{4}
\ArrowLine(50,20)(50,80)
\ArrowLine(50,80)(60,72.5)
\ArrowLine(80,57.5)(90,50)
\ArrowLine(90,50)(50,20)
\GlueArc(70,65)(12.5,142,322){3}{5}
\CArc(70,65)(12.5,142,322)
\DashCArc(70,65)(12.5,322,502){5}
\DashLine(90,50)(130,50){5}
\put(0,18){$g$}
\put(0,78){$g$}
\put(75,80){$\tilde Q$}
\put(60,40){$\tilde g$}
\put(70,25){$Q$}
\put(120,36){$A$}
\end{picture}
\begin{picture}(130,90)(0,0)
\Gluon(10,20)(50,20){-3}{4}
\Gluon(10,80)(50,80){3}{4}
\Line(50,80)(70,65)
\Gluon(50,80)(70,65){3}{3}
\Line(50,50)(50,80)
\Gluon(50,50)(50,80){3}{3}
\DashLine(70,65)(50,50){5}
\ArrowLine(70,65)(90,50)
\ArrowLine(90,50)(50,20)
\ArrowLine(50,20)(50,50)
\DashLine(90,50)(130,50){5}
\put(0,18){$g$}
\put(0,78){$g$}
\put(35,60){$\tilde g$}
\put(60,46){$\tilde Q$}
\put(80,60){$Q$}
\put(120,36){$A$}
\end{picture}
\begin{picture}(130,90)(-15,0)
\Gluon(10,80)(30,80){3}{2}
\Gluon(10,20)(50,20){-3}{4}
\Gluon(30,80)(50,80){3}{2}
\Line(30,80)(50,80)
\GlueArc(105.5,107.5)(80,200,244){3}{7}
\CArc(105.5,107.5)(80,200,244)
\CBoxc(51,47.5)(7,8){White}{White}
\DashLine(50,20)(50,80){5}
\ArrowLine(50,80)(90,50)
\ArrowLine(90,50)(70,35)
\DashLine(70,35)(50,20){5}
\DashLine(90,50)(130,50){5}
\put(0,18){$g$}
\put(0,78){$g$}
\put(28,53){$\tilde g$}
\put(57,12){$\tilde Q$}
\put(80,60){$Q$}
\put(120,36){$A$}
\end{picture} \\
\begin{picture}(130,90)(15,0)
\Gluon(10,80)(30,80){3}{2}
\Gluon(10,20)(50,20){-3}{4}
\Gluon(30,80)(50,80){3}{2}
\Line(30,80)(50,80)
\GlueArc(105.5,107.5)(80,200,244){3}{7}
\CArc(105.5,107.5)(80,200,244)
\CBoxc(51,47.5)(7,8){White}{White}
\ArrowLine(50,20)(50,80)
\DashLine(50,80)(90,50){5}
\DashLine(90,50)(70,35){5}
\ArrowLine(70,35)(50,20)
\DashLine(90,50)(130,50){5}
\put(0,18){$g$}
\put(0,78){$g$}
\put(28,53){$\tilde g$}
\put(57,12){$Q$}
\put(80,60){$\tilde Q$}
\put(120,36){$A$}
\end{picture}
\begin{picture}(130,90)(0,0)
\Gluon(10,20)(50,20){-3}{4}
\Gluon(10,80)(50,80){3}{4}
\Line(50,20)(50,80)
\Gluon(50,20)(50,80){3}{5}
\Line(50,80)(70,65)
\Gluon(50,80)(70,65){3}{2}
\Line(70,35)(50,20)
\Gluon(70,35)(50,20){3}{2}
\ArrowLine(70,65)(90,50)
\ArrowLine(90,50)(70,35)
\DashLine(70,65)(70,35){5}
\DashLine(90,50)(130,50){5}
\put(0,18){$g$}
\put(0,78){$g$}
\put(35,46){$\tilde g$}
\put(60,46){$\tilde Q$}
\put(80,60){$Q$}
\put(120,36){$A$}
\end{picture}
\begin{picture}(130,90)(-15,0)
\Gluon(10,20)(50,20){-3}{4}
\Gluon(10,80)(50,80){3}{4}
\Line(50,20)(50,80)
\Gluon(50,20)(50,80){3}{5}
\Line(50,80)(70,65)
\Gluon(50,80)(70,65){3}{2}
\Line(70,35)(50,20)
\Gluon(70,35)(50,20){3}{2}
\DashLine(70,65)(90,50){5}
\DashLine(90,50)(70,35){5}
\ArrowLine(70,65)(70,35)
\DashLine(90,50)(130,50){5}
\put(0,18){$g$}
\put(0,78){$g$}
\put(35,46){$\tilde g$}
\put(57,46){$Q$}
\put(80,60){$\tilde Q$}
\put(120,36){$A$}
\end{picture} \\
\caption{\label{fg:nlodia} \it Non-vanishing diagrams contributing to
the genuine SUSY--QCD corrections to pseudoscalar MSSM Higgs boson
production via gluon fusion mediated by top- and bottom quark ($Q=t,b$)
as well as stop/sbottom ($\tilde Q = \tilde t, \tilde b$) and gluino
($\tilde g$) loops at NLO.}
\end{center}
\end{figure}
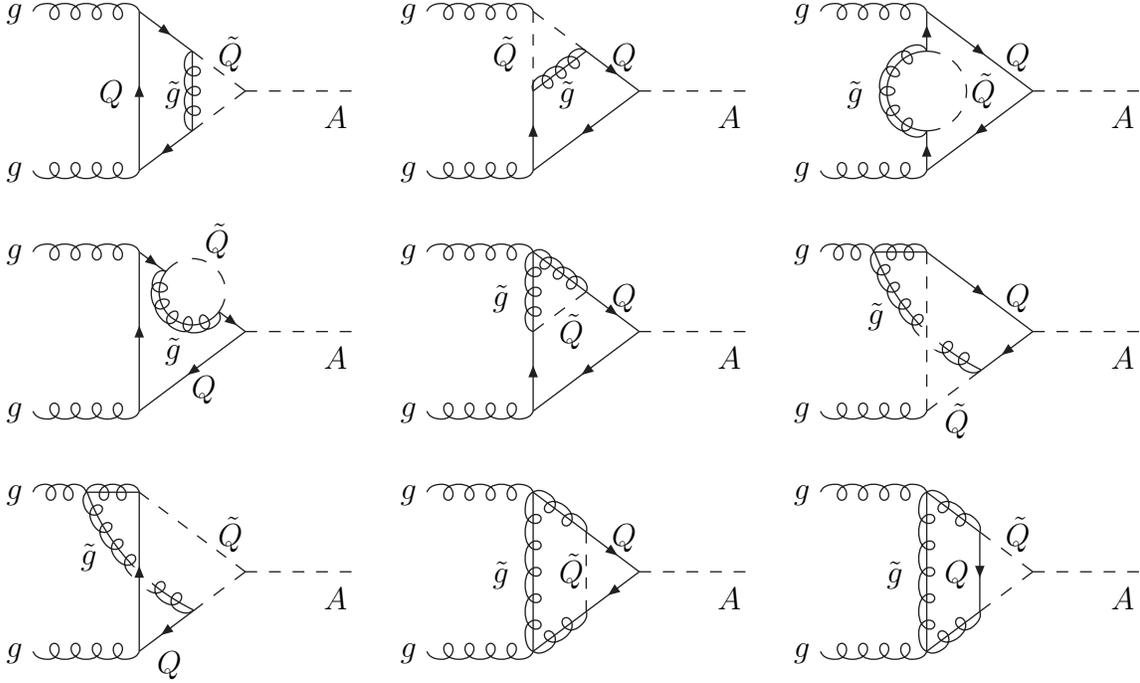
The genuine SUSY--QCD corrections at NLO are determined by the Feynman
diagrams shown in Fig.~\ref{fg:nlodia} that displays only the
non-vanishing graphs. Additional permutations of the external gluons
have to be added. The matrix element for the LO expression and the SUSY--QCD
corrections can be parametrized as
\begin{eqnarray}
{\cal M} & = & i\delta_{ab} \frac{\alpha_s}{2\pi v} {\cal T}^{\mu\nu}
\epsilon_\mu(q_1) \epsilon_\nu(q_2) \nonumber \\
{\cal T}^{\mu\nu} & = & A_{LO/SQCD}^A~\epsilon^{\mu\nu\alpha\beta} q_{1\alpha}
q_{2\beta} \;,
\end{eqnarray}
where $q_{1},q_{2}$ denote the two incoming momenta of the gluons and
$\epsilon_\mu(q_i)$ their polarization vectors,
$\delta_{ab}$ the Kronecker symbol of the adjoint $SU(3)_c$ color space
and $\epsilon^{\mu\nu\alpha\beta}$ the four-dimensional Levi--Civita
tensor. In this paper we will mainly use the $\gamma_5$ prescription of
Larin \cite{larin}, where the product of Levi--Civita tensors is
replaced by the determinant of $n$-dimensional metric tensors in
$n=4-2\epsilon$ dimensions. Within this framework we can construct a
projector on the anticipated form factors $A_{LO/SQCD}^A$,
\begin{equation}
{\cal P}^{\mu\nu} = \frac{2}{M_A^4 (1-\epsilon) (1-2\epsilon)}~
\epsilon^{\mu\nu\alpha\beta} q_{1\alpha} q_{2\beta}
\label{eq:projector}
\end{equation}
so that
\begin{equation}
{\cal P}^{\mu\nu} {\cal T}_{\mu\nu} = A_{LO/SQCD}^A \;.
\end{equation}
In order to set up a simple notation in close connection to the QCD
corrections of Eq.~(\ref{eq:mssmgghqcd5}) we will normalize the genuine
SUSY--QCD corrections to the individual form factors at LO,
\begin{equation}
A_{Q,SQCD}^A = A_{Q,LO}^A~{\cal C}_{Q,SQCD}^A~\frac{\alpha_s}{\pi} \;,
\label{eq:c_amp}
\end{equation}
where ${\cal C}_{Q,SQCD}^A$ depends on all ratios of the pseudoscalar Higgs,
quark, squark and gluino masses. The LO form factor
\begin{equation}
A_{Q,LO}^A = \Gamma(1+\epsilon) \left(\frac{4\pi \bar\mu^2}{m_Q^2}
\right)^\epsilon m_Q g_Q^A A_Q^A (\tau_Q)
\end{equation}
has been defined in terms of the expressions of
Eq.~(\ref{eq:loff}). In the following we will describe the technical
details for the numerical integration to determine the complex
coefficient ${\cal C}_{Q,SQCD}$ by exemplifying our method for the first
diagram of Fig.~\ref{fg:nlodia}. In order to regularize virtual
thresholds we have added a small imaginary part to the quark and squark
masses\footnote{This procedure is equivalent to adding an imaginary part
to the gluino mass in addition, but in our numerical analysis we do not
cross virtual thresholds involving the gluino so that this addition is
not required.},
\begin{equation}
m_Q^2 \to m_Q^2 (1-i\bar\epsilon), \qquad m_{\tilde Q_k}^2 \to m_{\tilde
Q_k}^2 (1-i\bar\epsilon) \qquad (k=1,2)\,, 
\label{eq:impart}
\end{equation}
with a positive regulator $\bar\epsilon > 0$, which defines the
analytical continuation of our two-loop amplitudes. We work with a small
but finite value of $\bar\epsilon$ that is small enough to achieve
results in the narrow-width approximation. For the parametrization of
the two-loop diagrams, we follow the same procedure used and described in
Refs.~\cite{hpair} for Higgs-boson pair production and adopted in
earlier works \cite{earlier}.

\subsection{Feynman Parametrization}
The parametrization of the first two-loop diagram of
Fig.~\ref{fg:nlodia} reads
\begin{eqnarray}
{\cal T}_1^{\mu\nu} & = & -\frac{C_F}{4}~\frac{g^A_{\tilde Q_l \tilde
Q_m}}{m_Q}~\frac{\alpha_s}{\pi}~(4\pi)^4~A_{1,lm}^{\mu\nu} \nonumber \\
A_{1,lm}^{\mu\nu} & = & \int \frac{d^n k d^n q}{(2\pi)^{2n}}
\frac{\mbox{Tr}\Big\{ \overline{I\!\!P}_l (-\slsh{q}+M_{\tilde g})
I\!\!P_m (\,\slsh{k}+\slsh{q}_1+m_Q) \gamma^\mu (\,\slsh{k}+m_Q)
\gamma^\nu (\,\slsh{k}-\slsh{q}_2+m_Q) \Big\}}{(k^2-m_Q^2)
[(k+q_1)^2-m_Q^2] [(k-q_2)^2-m_Q^2] [(k+q+q_1)^2-m_{\tilde Q_m}^2]}
\nonumber \\
& & \hspace*{2cm} \times \frac{1}{[(k+q-q_2)^2-m_{\tilde Q_l}^2]
(q^2 - M_{\tilde g}^2)}
\label{eq:loop}
\end{eqnarray}
where we sum over
  $l,m \in \{1,2\}$ in ${\cal T}_1^{\mu\nu}$, $k,q$ are the loop momenta
that are integrated over and the chiral coupling factors $I\!\!P_j$
$(j=1,2)$ are defined as
\begin{eqnarray}
I\!\!P_1 & = & I\!\!P_L \cos \theta_Q - I\!\!P_R \sin \theta_Q \nonumber \\
I\!\!P_2 & = & -I\!\!P_L \sin \theta_Q - I\!\!P_R \cos \theta_Q \nonumber \\
I\!\!P_{R/L} & = & \frac{1\pm\gamma_5}{2} \;,
\end{eqnarray}
and $\overline{I\!\!P}_j$ emerges from $I\!\!P_j$ by the replacement
$\gamma_5\to -\gamma_5$. After applying the contraction with the
projector ${\cal P}_{\mu\nu}$ onto the contribution to the virtual form
factor, we introduce Feynman parameters $x_3,x_4,x_1,x_2$ for the second
to fifth propagator (in this ordering) and $1-\sum_j x_j$ for the first
one, $(k^2-m_Q^2)$. With the substitutions
\begin{equation}
x_1 = (1-x)y,\quad x_2=(1-x)(1-y),\quad x_3 = x(1-z),\quad x_4 = xzv
\end{equation}
we obtain a four-dimensional Feynman-parameter integral over $x,y,z,v$
with integration boundaries from 0 to 1. The shift
\begin{eqnarray}
k & \to & k-Q_1 \nonumber \\
Q_1 & = & (1-x) q + [x+y-x(y+z)] q_1 - [(1-x)(1-y)+xzv] q_2
\end{eqnarray}
in both the numerator and denominator symmetrizes the $k$-integration
that is performed in a simple and systematic way for the emerging
integral. The residual $q$-dependent denominator after the
$k$-integration is treated as a propagator for the $q$-integration after
extracting all coefficients in front of the term $q^2$. We introduce a
fifth Feynman parameter $r$ for this propagator and $1-r$ for the
last purely $q^2$-dependent propagator of Eq.~(\ref{eq:loop}). Applying
the second shift
\begin{eqnarray}
q & \to & q-Q_2, \nonumber \\
Q_2 & = & -r(1-y-z) q_1 - r(1-y-zv) q_2
\end{eqnarray}
both in the numerator and denominator we perform the symmetric
$q$-integration. In this way, we finally arrive at an integral of the
type
\begin{equation}
A_{1,SQCD}^A = \frac{C_F}{4} \frac{g^A_{\tilde Q_l \tilde
Q_m}}{m_Q}~\frac{\alpha_s}{\pi}~\Gamma(2+2\epsilon)
\left(\frac{4\pi\mu_0^2}{M_{\tilde g}^2}\right)^{2\epsilon} \int_0^1 d^5 x~
\frac{x^{1+\epsilon}(1-x)^\epsilon z r^{2+\epsilon} H(\vec
x)}{N^{2+2\epsilon}(\vec x)} \;,
\label{eq:a1}
\end{equation}
with $\vec x = (x,y,z,v,r)$ and $d^5x = dx\,dy\,dz\,dv\,dr$. The term
$H(\vec x)$ denotes the full numerator and includes singular and higher
powers of the dimensional regulator $\epsilon$. $N(\vec x)$ is the final
denominator,
\begin{eqnarray}
N(\vec x) & = & x(1-x)(1-r) + \rho_Q xr + \rho_m (1-x)yr + \rho_l
(1-x)(1-y)r \nonumber \\
& + & \rho_A r\Big\{ x(1-x)r(1-y-z)(1-y-zv) \nonumber \\
& & \qquad - [y(1-x)+x(1-z)][(1-x)(1-y)+xzv] \Big\} \label{denominator}
\end{eqnarray}
where the ratios are defined as $\rho_Q = m_Q^2/M_{\tilde g}^2$,
$\rho_k=m_{\tilde Q_k}^2/M_{\tilde g}^2$, $\rho_A = M_A^2/M_{\tilde g}^2$.
This denominator is maximally a second-order polynomial in all Feynman
parameters we have introduced. The poles of $H(\vec x)$ in $\epsilon$
originate from powers of $k^2$ and $q^2$ in the numerators of the $k$-
and $q$-integrals. We have chosen the convention to normalize all mass
parameters to the gluino mass $M_{\tilde g}$. In order to cope with the
LO form factor in an easier way, we have rewritten the coefficients of
all integrals as
\begin{equation}
\Gamma(2+2\epsilon)
\left(\frac{4\pi\mu_0^2}{M_{\tilde g}^2}\right)^{2\epsilon} =
\Gamma^2(1+\epsilon) \left(\frac{4\pi\mu_0^2}{m_Q^2}\right)^\epsilon
\left(\frac{4\pi\mu_0^2}{M_{\tilde g}^2}\right)^\epsilon \times
\rho_Q^\epsilon (1+2\epsilon)(1+\epsilon^2 \zeta_2) + {\cal
O}(\epsilon^3).
\end{equation}
The factors $\rho_Q^\epsilon (1+2\epsilon)(1+\epsilon^2 \zeta_2)$ are
added to the integrands before expansion in $\epsilon$. The final
contribution to the coefficient ${\cal C}_{Q,SQCD}^A$ is then given by
\begin{eqnarray}
{\cal C}_{Q,SQCD}^{A, (1)} & = & \frac{C_F}{4} \frac{g^A_{\tilde Q_l \tilde
Q_m}}{m_Q^2 g_Q^A
A_Q^A(\tau_Q)}~\Gamma(1+\epsilon) \left(\frac{4\pi\mu_0^2}{M_{\tilde
g}^2}\right)^\epsilon \int_0^1 d^5 x~ \frac{x^{1+\epsilon}(1-x)^\epsilon
z r^{2+\epsilon} H(\vec x)}{N^{2+2\epsilon}(\vec x)} \nonumber \\
& & \hspace*{3cm} \times \rho_Q^\epsilon (1+2\epsilon+\epsilon^2
    \zeta_2) \;.
\end{eqnarray}
The final integral is finite for this diagram. For the other diagrams,
we follow the same procedure accordingly. All diagrams are infrared
finite, since all virtual particles are massive, but the residual
Feynman integrals contain end-point singularities in several cases that
are subtracted in the usual way according to the description of
Ref.~\cite{hpair}. The integration of the subtracted part yields the
corresponding UV singularities.

\subsection{Integration by Parts}
In our numerical analysis, we cross the virtual $b\bar b, t\bar t$
thresholds and for large pseudoscalar masses the $\tilde b_1 \tilde
b_2^\ast, \tilde t_1 \tilde t_2^\ast$ thresholds as well. The
parametrization of the integrals discussed so far is not sufficiently
stable above these thresholds due to the high power of the denominator
$N(\vec x)$ that becomes small in the Feynman-parameter regions in the
vicinity of the virtual thresholds. We need to adopt imaginary
regulators $\bar\epsilon \lsim 10^{-3}$ in order to obtain numbers
independent of this regulator. The small size of this regulator makes
the integral numerically unstable. A stabilization of the integration can be achieved by an integration by parts (IBP) to reduce the power of the denominator. In general, for this purpose, one can write
    \begin{align}
        \Delta &= p_0 N + \sum_i p_i \pdv{N}{x_i}
    \end{align}
    where $N$ is the dominator of the integral, $p_0$ and $p_i$ are polynomials and $\Delta$ is constant in the variables $x_i$. For simplicity we drop the arguments $\vec{x}$ everywhere. The polynomials $p_0$ and $p_i$ can be found by constructing the Gr\"obner basis of the set $\{N, \pdv{N}{x_i}\}$. We find that
    \begin{align}
        \int_0^1 d^n x \frac{g_m}{N^{m+ 2 \varepsilon}} &= \int_0^1 d^n x \frac{g_{m - 1}}{N^{m - 1+ 2 \varepsilon}} + \sum_i \qty[\frac{g_{m - 1}^{(i)}}{N^{m - 1+ 2 \varepsilon}}]_{x_i = 0}^{x_i = 1} \nonumber \\
        g_{m - 1} &= \frac{1}{\Delta} \qty(g_m p_0 - \frac{\sum_i \partial_{x_i} (g_m p_i)}{1 - m - 2 \varepsilon} ) \nonumber \\
        g_{m - 1}^{(i)} &= \frac{1}{\Delta} \frac{g_m p_i}{1 - m - 2 \varepsilon}.
    \label{eq:ibp}
    \end{align}
    These equations can be applied iteratively to reduce the power of the denominator further.\\
     Not every choice of the parameters for the integration by parts will yield a stable result. Potential issues can arise from singularities in the boundary terms as well as singularities that arise when $\Delta = 0$, which can happen when $N = 0$ and all $\pdv{N}{x_i} = 0$.\\
     Choosing only a subset of the Feynman parameters yields shorter expressions that can be evaluated faster. For practical purposes, it is thus usually best to find a parametrization where using a single Feynman parameter for the integration by parts is sufficient to stabilize the numerical integration. \\
We exemplify the two examples encountered in our calculation. If
 $N$ is linear in the Feynman parameter $x_1$, i.e.
     \begin{align}
         N = a x_1 + b,
     \end{align}
     there are two possible choices for the polynomials
     \begin{align}
         p_0 &= 0 & p_1 &= 1 & \Delta &= a \\
         \mathrm{or}\quad p_0 &= 1 & p_1 &= -x_1 & \Delta &= b.
     \end{align}
     A linear combination of these two solutions is also valid. If $N$ is quadratic in the Feynman parameter i.e.
     \begin{align}
        N = a x_1 ^2 + b x_1 + c,
    \end{align}
    the polynomials are given by
    \begin{align}
        p_0 &= 4 a & p_1 &= - b - 2 a x_1 & \Delta &= -b^2 + 4 a c.
        \label{eq:Delta}
    \end{align}
    For example in the first diagram we have achieved stabilization for $x_1 = v$. The denominator is linear in this parameter and we have
\begin{eqnarray}
N(\vec x) & = & av + b \nonumber \\
a & = & -\rho_A x z r\Big\{ r(1-z) + (1-r) [y(1-x)+x(1-z)] \Big\}
\nonumber \\
b & = & x(1-x)(1-r) + \rho_Q xr + \rho_m (1-x)yr + \rho_l
(1-x)(1-y)r \nonumber \\
& + & \rho_A (1-x) (1-y) r\Big\{ x r (1-y-z) - [y(1-x)+x(1-z)] \Big\} \;.
\end{eqnarray}
With this explicit parametrization at
  hand, the following manipulation can be performed,
\begin{eqnarray}
\int_0^1 dv~\frac{H_i(\vec x)}{N^2(\vec x)} & = &
\frac{H(\vec x)|_{v=0}}{ab} - \frac{H(\vec x)|_{v=1}}{a(a+b)}
+ \frac{[\partial_v H(\vec x)]|_{v=1}\log(a+b) - [\partial_v H(\vec
x)]|_{v=0}\log(b)}{a^2} \nonumber \\
& & - \int_0^1 \frac{dv}{a^2}~[\partial_v^2 H(\vec x)] \log(av+b) \;,
\end{eqnarray}
according to Eq.~(\ref{eq:ibp}). Since the powers of all denominators are reduced and the original
denominator $N(\vec x)$ appears in the argument of a logarithm in the
last integral the numerical integration appears to be stable for the
imaginary regulator down to $\bar\epsilon\lsim 10^{-4}$ which is
sufficient for the narrow-width limit.

In cases of a Feynman parameter entering the
denominator in second order,
\begin{equation}
N(\vec x) = ay^2 + by + c
\end{equation}
and making make use of the identities of Eq.~(\ref{eq:Delta}) (we drop the arguments of $N$)
\begin{equation}
\Delta = 4 ac - b^2 = 4aN - (\partial_y N)^2 = 4aN - (2ay+b)^2 \;,
\end{equation}
we arrive at the special situation that the derivative appears in second power. This allows us to
perform two IBPs of the original integral \cite{hpair},
\begin{equation}
\int_0^1 dy~\frac{H}{N^2} = \frac{1}{\Delta} \left\{
\left. \left[ \frac{2ay+b}{N}H -(\partial_y H) \log N \right]
\right|^{y=1}_{y=0} + \int_0^1 dy~\left[ \frac{2a}{N} H +
(\partial_y^2 H) \log N \right] \right\},
\end{equation}
where for simplicity we dropped the arguments $\vec x$ everywhere.

\subsection{Renormalization}
In our calculation of the genuine SUSY--QCD corrections we have to
renormalize the SUSY--QCD part of the quark mass only, since everything
else is already accounted for by the QCD corrections, i.e.~the decoupling of
all SUSY particles from the evolution of the strong coupling $\alpha_s$
and the PDFs that both run with five active flavours in our calculation.
The SUSY--QCD part of the on-shell quark-mass counterterm is given by
[see Eq.~(\ref{eq:dmq})]
\begin{eqnarray}
\frac{\delta m_Q}{m_Q} & = & -C_F\frac{\alpha_s}{4\pi} \left\{
B_1(m_Q^2;M_\sgl,m_{\tilde Q_1}) + B_1(m_Q^2;M_\sgl,m_{\tilde Q_2})
\right.  \nonumber \\
& & \left. \hspace*{0.7cm} +2 M_\sgl (A_Q-\mu r_Q)
\frac{B_0( m_Q^2;M_\sgl,m_{\tilde Q_1}) - B_0(m_Q^2;M_\sgl,m_{\tilde
Q_2})}{m^2_{\tilde Q_1}-m^2_{\tilde Q_2}}\right\} \, .
\end{eqnarray}
We renormalize the quark mass on-shell, because the LO form
factor $A_Q^A(\tau_Q)$ and the pure QCD corrections are expressed in
terms of the quark pole mass\footnote{For the evaluation of the NLO SUSY--QCD contributions, however, we use the derived bottom mass $\hat m_b$ [see Eq.(\ref{eq:derivedbotmass})] in the calculation of $C_{b,SQCD}^A$. The resulting difference only contributes at the NNLO level.}. The corresponding counterterm for the
gluon-fusion cross section form factor is given by
\begin{equation}
\delta_1 A_{Q,SQCD}^A = \frac{\partial \tilde A_Q^A(\tau_Q)}{dm_Q} \delta m_Q =
2~\tau_Q\frac{\partial \tilde A_Q^A(\tau_Q)}{\partial\tau_Q}~ \frac{\delta
m_Q}{m_Q} \;,
\end{equation}
where $\tilde A_Q^A(\tau_Q)$ denotes the LO form factor including ${\cal
O}(\epsilon)$ terms,
\begin{eqnarray}
\tilde A_Q^A(\tau) & = & \tau f(\tau) + \epsilon \frac{\tau}{4} H(\tau) +
{\cal O}(\epsilon^2) \nonumber \\
H(\tau) & = & 4\left\{ S_{1,2}(x) + S_{1,2}\left(\frac {1}{x}\right)
\right\} + 2\left\{ Li_3(x) + Li_3\left(\frac {1}{x}\right) \right\} +
2\zeta_3 \nonumber \\
x & = & \frac{1-\sqrt{1-\tau}}{1+\sqrt{1-\tau}} \;,
\label{eq:aqlon}
\end{eqnarray}
with the usual trilogarithms,
\begin{eqnarray}
S_{1,2}(y) & = & \frac{1}{2} \int_0^1 \frac{dz}{z} \log^2(1-zy)
\nonumber \\
Li_3(y) & = & \int_0^1 \frac{dz}{z} \log(z) \log(1-zy) \;.
\end{eqnarray}
The derivative is given by
\begin{eqnarray}
\tau\frac{\partial \tilde A_Q^A(\tau)}{\partial\tau} & = & \tilde
A_Q^A(\tau) + \frac{\tau}{1-\tau} g(\tau) + \frac{\epsilon}{2}\left\{
\frac{\tau}{\tau-1} g(\tau) \log\left( 4\frac{\tau-1}{\tau}\right)
\right. \nonumber \\
& + & \left. \frac{\tau}{\sqrt{1-\tau}} \left[ Li_2\left(
\frac{1}{1-x}\right) - Li_2\left( \frac{-x}{1-x}\right) \right] \right\}
\nonumber \\[0.5cm]
g(\tau) & = & \left\{ \begin{array}{ll} \displaystyle \sqrt{\tau-1} \arcsin
\frac{1}{\sqrt{\tau}} & \tau \ge 1 \\
\displaystyle \frac{\sqrt{1-\tau}}{2} \left[ \log \frac{1+\sqrt{1-\tau}}
{1-\sqrt{1-\tau}} - i\pi \right] & \tau < 1 \end{array} \right.
\end{eqnarray}
where $Li_2$ denotes the dilogarithm,
\begin{equation}
Li_2(y) = -\int_0^1 \frac{dz}{z} \log(1-zy) \; .
\end{equation}
However, we introduce effective low-energy Yukawa couplings in our
calculation, i.e.~the Yukawa couplings of a low-energy Two-Higgs-Doublet
model (2HDM), where the heavy SUSY particles are integrated out. This
implies that the top- and bottom-Yukawa couplings are dressed with
$\Delta_{t/b}$ contributions. The SUSY--QCD parts of these contributions
are given by
\begin{eqnarray}
\Delta_Q & = & \frac{C_F}{2}~\frac{\alpha_s(\mu_R)}{\pi}~M_{\sgl}~\mu~
r_Q~ I(m^2_{\tilde Q_1},m^2_{\tilde Q_2},M^2_{\sgl}) \nonumber \\
I(a,b,c) & = & \frac{\displaystyle ab\log\frac{a}{b} + bc\log\frac{b}{c}
+ ca\log\frac{c}{a}}{(a-b)(b-c)(a-c)} \;.
\label{eq:deltaq}
\end{eqnarray}
The expressions for the Yukawa couplings including resummations of the
leading $\cot\beta$-enhanced contributions for the top-Yukawa coupling
and the $\tgb$-enhanced terms of the bottom Yukawa coupling can be cast
into the form
\begin{eqnarray}
g_Q^A \to \tilde g^A_Q & = & \frac{g^A_Q}{1+\Delta_Q}\left[ 1 -
\frac{\Delta_Q}{r_Q^2} \right] \;,
\label{eq:rescoup}
\end{eqnarray}
with $r_Q$ defined after Eq.~(\ref{eq:matrix}).  These contributions
will result in additional terms in the counterterms of our calculation,
\begin{eqnarray}
\Delta A_{Q,SQCD}^A & = & A_Q^A(\tau_Q) \left( 1+\frac{1}{r_Q^2}
\right) \Delta_Q \, ,
\label{eq:ctdel}
\end{eqnarray}
since the LO form factors $A_Q^A(\tau_Q)$ are proportional to the {\it
linear} quark-Yukawa coupling. This results in the complete counterterm
\begin{equation}
\delta A_{Q,SQCD}^A = 2~\tau_Q\frac{\partial \tilde
A_Q^A(\tau_Q)}{\partial\tau_Q}~ \frac{\delta m_Q}{m_Q} + \Delta A_Q^A(\tau_Q)
\end{equation}

\subsection{Hadronic Cross Section}
Our notation can be viewed as a modification of the factor $\sigma^A_0$ of
Eq.~(\ref{eq:locxn}) as a starting point that can easily be extended
to the NLO corrections,
\begin{eqnarray}
\sigma^A_0 & = & \frac{G_F \alpha_s^2}{128\sqrt{2} \pi} \left| g_t^A
A_t(\tau_t) \left(1+\hat {\cal C}^A_{t,SQCD} \frac{\alpha_s}{\pi} \right) + g_b^A
A_b(\tau_b) \left(1+\hat {\cal C}^A_{b,SQCD} \frac{\alpha_s}{\pi} \right) \right|^2
\nonumber \\
& = & \frac{G_F \alpha_s^2}{128\sqrt{2} \pi} \left\{ \left| \tilde g_t^A
A_t(\tau_t) + \tilde g_b^A A_b(\tau_b) \right|^2 \right.
\label{eq:csqcd} \\
& + & \left. 2\Re \left[ \left[\tilde g^A_t A_t(\tau_t) + \tilde g^A_b
A_b(\tau_b)\right]^\ast \left[ g^A_t A_t(\tau_t) {\cal C}^A_{t,SQCD} + g^A_b
A_b(\tau_b) {\cal C}^A_{b,SQCD} \right] \frac{\alpha_s}{\pi} \right] + {\cal
O}(\alpha_s^2) \right\} \nonumber
\end{eqnarray}
where $\tilde g_Q^A$ ($Q=t,b$) denote the resummed quark Yukawa
couplings of Eq.~(\ref{eq:rescoup}) that absorb $\Delta_b$ and $\Delta_t$
contributions in the effective Yukawa couplings as the appropriate
effective Yukawa couplings in the low-energy effective 2HDM. The factors ${\cal C}^A_{Q,SQCD}$ and $\hat{\cal C}^A_{Q,SQCD}$ ($Q=t,b$) denote the relative SUSY--QCD corrections factors to the individual form factors with and without absorption of the $\Delta_Q$ terms, respectively. Within this
framework the Yukawa couplings of the QCD corrections will be replaced
by these effective Yukawa couplings as well due to the factorizing
properties of EFT couplings from the pure QCD corrections.
However, the subleading contributions of Eq.~(\ref{eq:csqcd}) involve
the LO Yukawa coulings, since $\Delta_{t,b}$ effects only factorize at
the leading order of an $1/M^2_{SUSY}$ expansion so that the SUSY--QCD remainder does not factorize from the effective Yukawa couplings in general. This will avoid
artificial singularities in the scalar MSSM Higgs sector as well
\cite{haefliger}. Expressing the LO factor $\sigma^A_0$ in terms of the
effective Yukawa couplings,
\begin{equation}
\sigma^A_0 \to \tilde \sigma^A_0 = \frac{G_F \alpha_s^2}{128\sqrt{2} \pi} \left| \tilde
g_t^A A_t(\tau_t) + \tilde g_b^A A_b(\tau_b) \right|^2 \;,
\end{equation}
and referring to Eq.~(\ref{eq:c_sqcd}), the SUSY--QCD corrections
add to the virtual coefficient $C^A$,
\begin{equation}
C^A = C_{QCD}^A + C_{SQCD}^A \;,
\end{equation}
with the usual QCD-correction coefficient $C_{QCD}^A$ and
\begin{equation}
C_{SQCD}^A = 2\Re \left\{ \frac{g^A_t A_t(\tau_t) {\cal C}^A_{t,SQCD} + g^A_b
A_b(\tau_b) {\cal C}^A_{b,SQCD}}{\tilde g^A_t A_t(\tau_t) + \tilde g^A_b
A_b(\tau_b)} \right\} \;,
\end{equation}
where we are using LO Yukawa couplings $g_Q^A$ in the numerator, since this
contribution constitutes the remainder of the full SUSY--QCD
corrections that does not factorize in general terms. In
Eq.~(\ref{eq:csqcd}) and for the following
discussion of the results, we distinguish between this coefficient for
the SUSY-remainder and the corresponding coefficient\footnote{Note that in the case of $\hat C^A_{Q,SQCD}$ we have to normalize to the LO expression with LO, i.e.~{\it without} effective, Yukawa couplings.},
\begin{eqnarray}
\hat C_{SQCD}^A & = & \overline{C}_{SQCD}^A - 2 \Re \left\{ \frac{g^A_t \Delta
A_{t,SQCD} + g^A_b \Delta A_{b,SQCD}}{g_t^A A_t(\tau_t) + g^A_b
A_b(\tau_b)} \right\} \nonumber \\
\overline{C}_{SQCD}^A & = & 2\Re \left\{ \frac{g^A_t A_t(\tau_t) {\cal C}^A_{t,SQCD} + g^A_b
A_b(\tau_b) {\cal C}^A_{b,SQCD}}{g^A_t A_t(\tau_t) + g^A_b A_b(\tau_b)} \right\}
\end{eqnarray}
that describes the full SUSY--QCD corrections without introducing the
effective top and bottom Yukawa couplings, i.e.~without absorbing
$\Delta_Q$ terms in the Yukawa couplings. The contributions $\Delta
A_{Q,SQCD}$ are given in Eq.~(\ref{eq:ctdel}).

\subsection{Axial $\gamma_5$ Schemes}
We have implemented the Larin scheme of Ref.~\cite{larin} that is a
variant of the original 't Hooft--Veltman scheme that has been set-up
systematically by Breitenlohner and Maison \cite{thoovel}. We have
extracted the Levi--Civita tensor at the pseudoscalar vertex by means of
the replacement
\begin{equation}
\gamma_5 = \frac{i}{24}~\epsilon_{\mu\nu\rho\sigma}
\gamma^\mu\gamma^\nu\gamma^\rho\gamma^\sigma \label{gamma5}
\end{equation}
and just keeping the four $\gamma$ matrices inside the traces. The
  diagrams where the pseudoscalar couples to squarks do not have such
  a vertex. They are however finite, such that a naively
  anticommuting $\gamma_5$ can be used at NLO. The
chiral couplings at the quark-squark-gluino vertices are treated fully
anticommuting to arrive at traces with one or no $\gamma_5$ matrix. Only
the contributions with no additional $\gamma_5$ matrix contribute after
applying the projector of Eq.~(\ref{eq:projector}). The projector yields
a product of two Levi--Civita tensors that is defined as
\begin{equation}
\epsilon^{\mu\nu\rho\sigma} \epsilon_{\mu'\nu'\rho'\sigma'} = -\mathrm{Det}
\left[ \begin{array}{cccc}
g^\mu_{\mu'} & g^\mu_{\nu'} & g^\mu_{\rho'} & g^\mu_{\sigma'} \\
g^\nu_{\mu'} & g^\nu_{\nu'} & g^\nu_{\rho'} & g^\nu_{\sigma'} \\
g^\rho_{\mu'} & g^\rho_{\nu'} & g^\rho_{\rho'} & g^\rho_{\sigma'} \\
g^\sigma_{\mu'} & g^\sigma_{\nu'} & g^\sigma_{\rho'} & g^\sigma_{\sigma'}
\end{array} \right] \;,
\end{equation}
where the metric tensors inside this determinant are treated as
$n$-dimensional objects. This prescription avoids a splitting of
$\gamma$ matrices and loop momenta into 4- and $(n-4)$-dimensional
components. Since $\gamma_5$ as defined in Eq.~(\ref{gamma5}) does not anticommute, an anomalous counterterm has to be added. However the genuine SUSY--QCD contributions to this counterterm vanish. In the 't Hooft--Veltman scheme, the metric tensors
in this determinant are defined as strictly 4-dimensional objects so
that the numerators of the loop integrals split into 4- and
$(n-4)$-dimensional pieces that have to be treated separately. To avoid additional anomalous counterterms we used anticommuting $\gamma_5$
matrices at the $Q\tilde Q\tilde g$-vertices in this scheme as well. We found full agreement for both
schemes. In addition, we have lifted the anti-commuting properties of
the $\gamma_5$ matrices entering at the $Q\tilde Q\tilde g$-vertices and
found mismatches that require anomalous subtractions to restore the
chiral properties. Finally, we have implemented the $\gamma_5$ scheme of
Ref.~\cite{kreimer} that gives up the cyclicity of the traces but keeps
the full anti-commuting property of the $\gamma_5$ matrix. The cyclicity of the trace is equivalent to the arbitrary decision where we start to read the fermion lines. To resolve this, the scheme defines unambiguous reading points in each diagram relative to the external axial couplings. However, since we have no axial vector couplings in our diagrams, the only prescription we have to follow is that the reading point must be outside of subdivergences, e.g. in the fifth diagram of Fig.~\ref{fg:nlodia} the reading point must not be at the $g\tilde g\tilde g$ vertex. We found full agreement with the calculation in the Larin scheme as well.

Finally, we have reproduced the limit of large top, stop and gluino
masses of Ref.~\cite{slavich} and found full agreement.
Ref.~\cite{slavich} worked with Pauli--Villars regularization so that
their Clifford algebra is defined in four dimensions strictly resulting in a fully
anti-commuting $\gamma_5$. That we have found full agreement
with this calculation in the large-mass limit underlines the consistency
of our results.

\subsection{Adler--Bardeen Theorem}
According to the analytical results of Ref.~\cite{slavich} the SUSY--QCD
coefficient in the large SUSY-mass limit (keeping the quark mass small)
is given by
\begin{equation}
\hat {\cal C}^A_{Q,SQCD} = -\frac{C_F}{2} \frac{M_{\tilde g}}{m_Q} \left(
\frac{s_{2\theta_Q}}{2} - \frac{m_Q Y_Q}{m^2_{\tilde Q_1} - m^2_{\tilde
Q_2}} \right) \left( \frac{\rho_1}{1-\rho_1} \log \rho_1 -
\frac{\rho_2}{1-\rho_2} \log \rho_2 \right) + {\cal O}(M_{SUSY}^{-2}),
\end{equation}
with $\rho_i$ as defined after Eq.~(\ref{denominator}). In this expression, we have focused
just on the leading terms of the large-mass expansion, since this is the
relevant contribution of the matching to a low-energy 2HDM. Moreover, in
the expression above the $\Delta_Q$ terms are not subtracted, i.e. this
is the result in terms of the LO Higgs coupling $g_Q^A$ without
$\Delta_Q$-dressing. The coupling $Y_Q$ is related to the squark
coupling,
\begin{equation}
Y_Q = 2 \frac{g_{\tilde Q_1 \tilde Q_2}}{m_Q g_Q^A} = A_Q +
\frac{\mu}{r_Q} \;.
\end{equation}
Inserting the explicit expressions for $s_{2\theta_Q}$ and $Y_Q$ one
arrives at
\begin{equation}
\hat {\cal C}^A_{Q,SQCD} = -\Delta_Q \left( 1+\frac{1}{r_Q^2} \right) + {\cal
O}(M_{SUSY}^{-2}) \;.
\end{equation}
Since the $At\bar t$
operator mixes with the $A\tilde t \tilde t^*$ operator the
non-decoupling $\Delta_t$ contributions to the effective top Yukawa
coupling are induced. Working with properly matched low-energy parameters, i.e.~effective
Yukawa couplings with $\Delta_Q$ contributions as in
Eq.~(\ref{eq:rescoup}), this term is absorbed in the Yukawa couplings
exactly so that the radiative corrections in the low-energy 2HDM with
properly defined low-energy parameters are vanishing for the leading
${\cal O}(M_{SUSY}^0)$ term
\begin{equation}
{\cal C}^A_{Q,SQCD} = {\cal O}(M_{SUSY}^{-2}).
\end{equation}
This is because in contrast to the MSSM, the chiral symmetry
  $\psi_Q \rightarrow e^{i \alpha \gamma_5} \psi_Q$ is only broken by
  the quark mass term in the effective 2HDM so that only the
    higher-order corrections to the proper matching of the low-energy
    2HDM to the full MSSM contribute.
Thus, in the low-energy limit the Adler-Bardeen theorem
\cite{adlerbardeen} is fulfilled\footnote{The Adler-Bardeen theorem is
{\it not} valid for subleading ${\cal O}(M_{SUSY}^{-2})$ orders in the
large SUSY-mass expansion with effective low-energy parameters as it
also not valid for subleading ${\cal O}(m_t^{-2})$ orders of the
large-top mass expansion of the pure QCD corrections.}. Since
radiative corrections still arise due to the higher-order corrections to
the matching, the Adler--Bardeen theorem \cite{adlerbardeen} builds a deep
connection between the explicit structure of the radiative corrections
in the full MSSM in the low-energy limit and the radiative corrections
in the low-energy EFT. This result is in line with the result of
Ref.~\cite{aggnnlo} that the QCD corrections to the effective $ggA$
Lagrangian in the HTL are vanishing if the strong coupling is chosen as
the 5-flavour one, i.e.~properly decoupling the top-quark contribution
from the running of $\alpha_s$ or in other words using the properly
matched low-energy $\alpha_s$ within pure 5-flavour QCD. This also
implies that in the large SUSY-mass limit (keeping the top mass small in
comparison) no effective $ggA$ operator is generated in the low-energy
2HDM at the dimension-5 level by integrating out the SUSY particles. The
same is true as well for the bottom/sbottom contributions so that the
SUSY particles do not generate a sbottom-induced effective $ggA$
operator at leading ${\cal O}(M^0_{SUSY})$ at all.

Another situation arises when the top quark is integrated out, i.e.~assumed
to be much heavier than the pseudoscalar $A$ as well and {\it not}
assumed to be much lighter than the other SUSY particles. In this case a
dimension-5 operator contribution is generated on top of the HTL at LO
due to the non-decoupling nature of the top quark as has already been
observed for the leading ${\cal O}(G_F m_t^2)$ corrections to the
effective $Agg$ coupling \cite{kniehl}. Since the stops couple to the pseudoscalar in terms of the
top Yukawa coupling as well, a new genuine dimension-5
contribution to the $Agg$ coupling, on top of the contribution from the effective Yukawa coupling, emerges starting at NLO. This
contribution can be related to the violation of the global Peccei--Quinn
symmetry of the MSSM Lagrangian by the $\mu$ term \cite{gganlosqcd,
slavich}. This leads to an extension of the
related operator identity of the divergence of the axial-vector current
by an additional operator involving the stop fields thus destroying the
one-to-one correspondence between the pseudoscalar top-Yukawa coupling
and the ABJ-anomaly operator and in this way the translation of the
Adler--Bardeen theorem to the $Agg$ operator. It follows that both the $\Delta_t$ terms and the genuine radiative corrections to the $Agg$ coupling scale with the $\mu$ parameter \cite{lukas}. Within the EFT view this has to be considered as
higher-order corrections to the effective $Agg$ operator in the combined
HTL and large-SUSY-mass limit, i.e.~higher-order corrections to the
corresponding matching conditions that scale with $\mu$.

\section{Results} \label{sc:results}
We are now in the position to present and discuss the final results of
the NLO SUSY--QCD corrections to pseudoscalar $gg\to A$ production, but
also to the pseudoscalar decays $A\to gg$ and $A\to \gamma\gamma$.
For the numerical analysis we have adopted the $M_h^{125}$ benchmark
scenario \cite{bench} that is defined by the following on-shell parameters,
\begin{eqnarray}
\mbox{$M_h^{125}$:} && M_{\tilde Q} = 1.5~{\rm TeV},\quad
M_{\tilde \ell_3} = 2~{\rm TeV},\quad M_{\sgl} = 2.5~{\rm TeV},
\nonumber \\
&& M_1 = M_2 = 1~{\rm TeV},\quad A_b = A_\tau = A_t = 2.8~{\rm
TeV} + \mu/\tgb,\quad \mu = 1~{\rm TeV},
\end{eqnarray}
that have been used in the framework of the program {\tt HDECAY}
\cite{hdecay} with an iteration to determine the corresponding
$\overline{\rm MS}$ parameters accordingly. This proceeds along the
lines discussed in Section \ref{sc:squark}. Here, $M_{\tilde Q}$ denotes
the third-generation soft SUSY-breaking squark-mass parameters,
$M_{\tilde \ell_3}$ the corresponding one for the sleptons and $M_1,
M_2$ the soft SUSY-breaking gaugino-mass parameters for the bino and
wino, repectively. For two representative values of $\tgb$, the related
stop and sbottom masses amount to
\begin{eqnarray}
&& \underline{\tgb=10} \nonumber \\
&& m_{\stopx_1} = 1340~{\rm GeV}, \quad m_{\stopx_2} =
1662~{\rm GeV}, \quad m_{\sbottom_1} = 1496~{\rm GeV}, \quad
m_{\sbottom_2} = 1508~{\rm GeV} \nonumber \\
&& \underline{\tgb=40} \nonumber \\
&& m_{\stopx_1} = 1340~{\rm GeV}, \quad m_{\stopx_2} =
1662~{\rm GeV}, \quad m_{\sbottom_1} = 1479~{\rm GeV}, \quad
m_{\sbottom_2} = 1525~{\rm GeV}.
\end{eqnarray}
Our numerical integration has been performed with the {\tt VEGAS} subroutine \cite{vegas} after preparing the integrands according to the methods described in Section \ref{sc:nlo}. We have used up to $\order{10^9}$ points for the 5-dimensional {\tt VEGAS} integration, with imaginary parts $\bar\epsilon$ of Eq.~(\ref{eq:impart}) up to the order of $10^{-3}$ above the virtual thresholds ($Q\bar Q, \tilde Q_1 \overline{\tilde Q}_2, \tilde Q_2 \overline{\tilde Q}_1)$ for $Q = t,b$. The numerical integration errors of our final results rank below the $10^{-2}$-level for the final coefficients ${\cal C}^A_{Q,SQCD}$ of Eq.~(\ref{eq:c_amp}) and ${\cal D}^A_{Q,SQCD}$ of Eq.~(\ref{eq:d_amp}) for both the top- and bottom-induced corrections. This has been achieved with less than a week of CPU time for each individual $M_A$ point.

\subsection{Gluon Fusion $gg\to A$}
\begin{figure}[hbtp]
\begin{center}
\begin{picture}(150,240)(0,0)
\put(-180,-70.0){\includegraphics{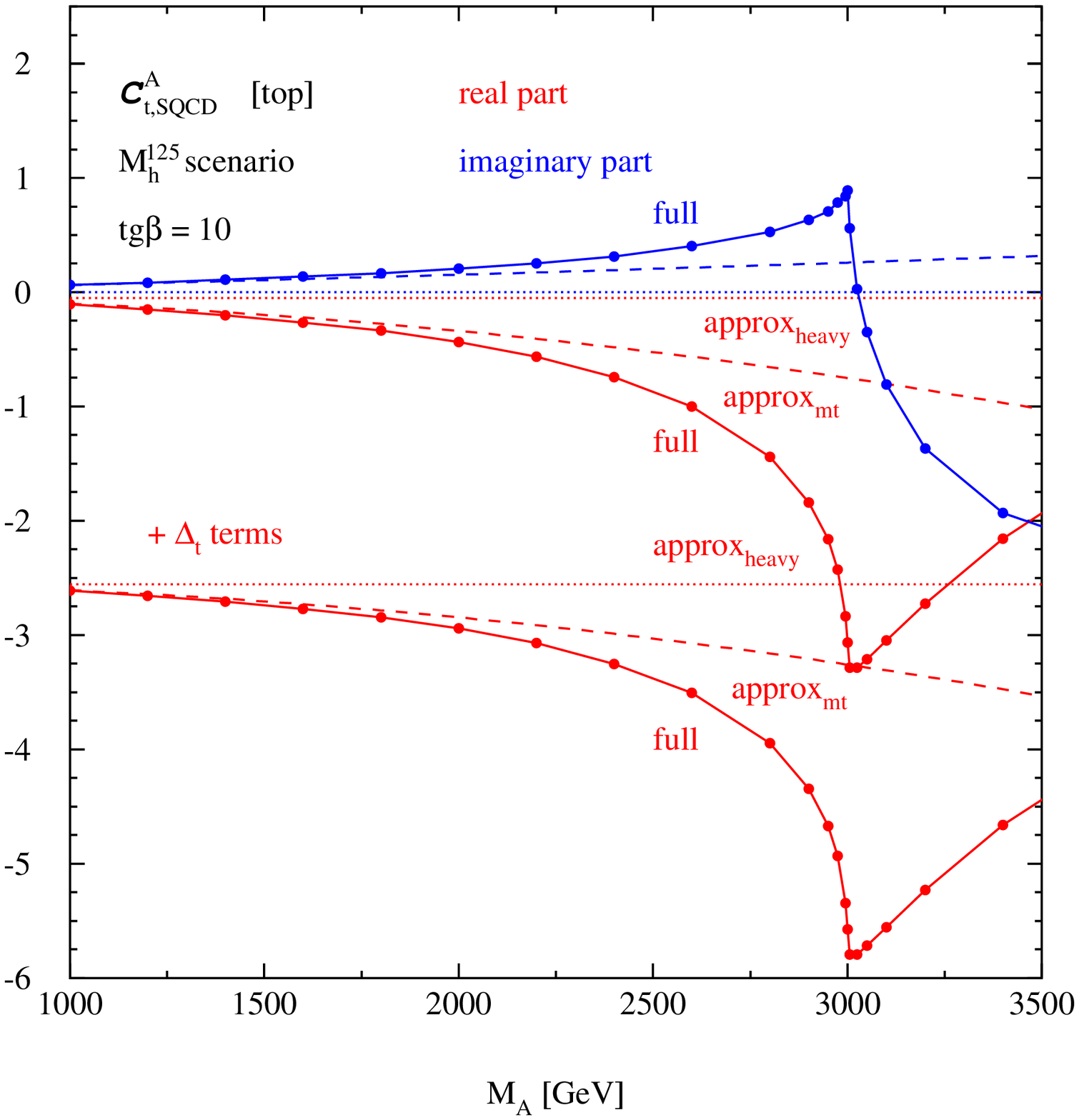}}
\put(60,-70.0){\includegraphics{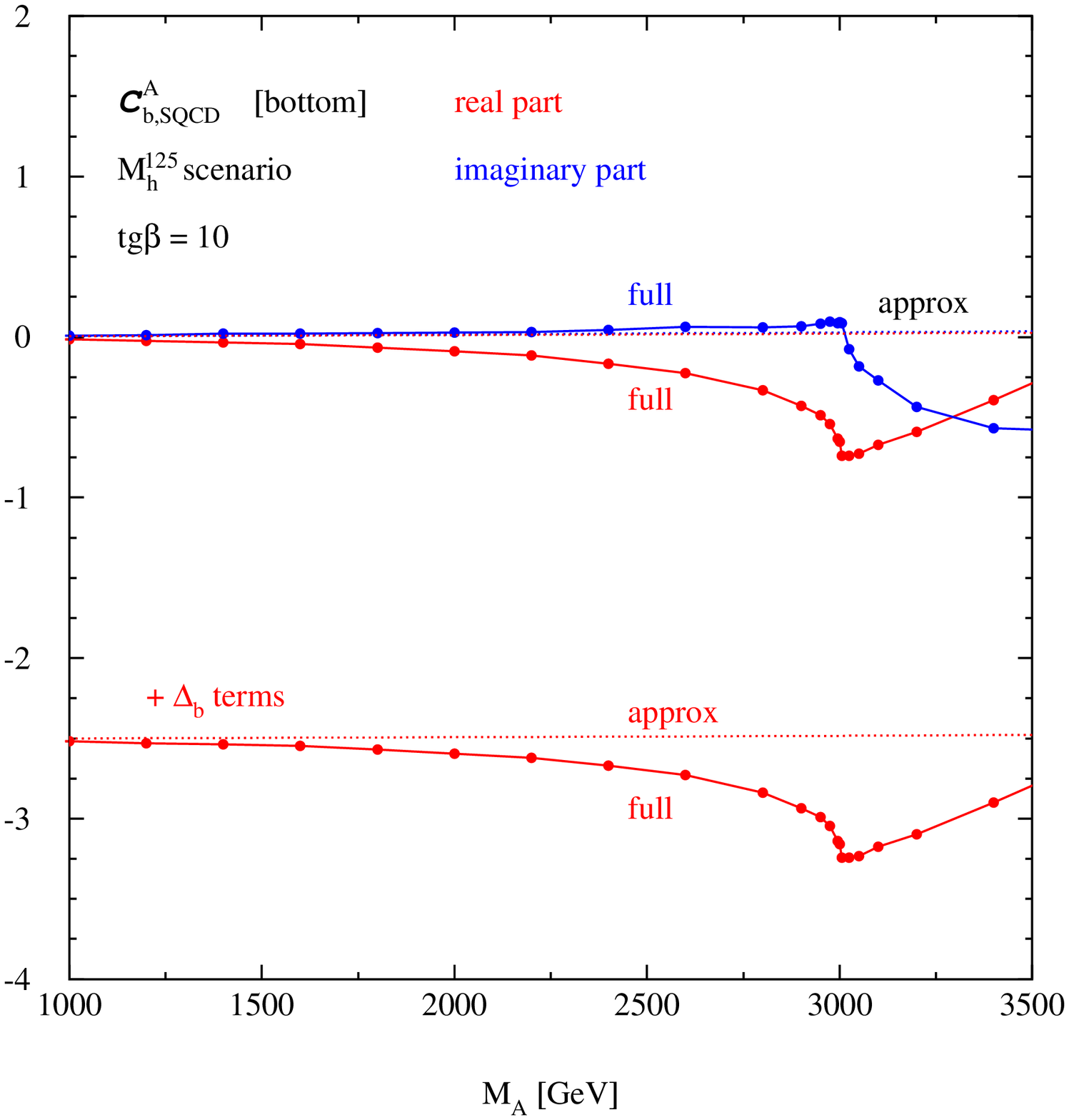}}
\end{picture}
\caption{\it The genuine SUSY--QCD corrections to $gg\to A$ normalized
to the LO top and bottom quark form factors for $\tgb=10$ in the
$M_h^{125}$ benchmark scenario. Real part: red, imaginary part: blue,
compared to the approximate calculations of Ref.~\cite{slavich} (dashed
lines). The dotted lines for the stop contributions correspond the the
combined limit of large top and SUSY masses.}
\label{fg:csusy10}
\end{center}
\end{figure}
\begin{figure}[hbtp]
\begin{center}
\begin{picture}(150,240)(0,0)
\put(-180,-70.0){\includegraphics{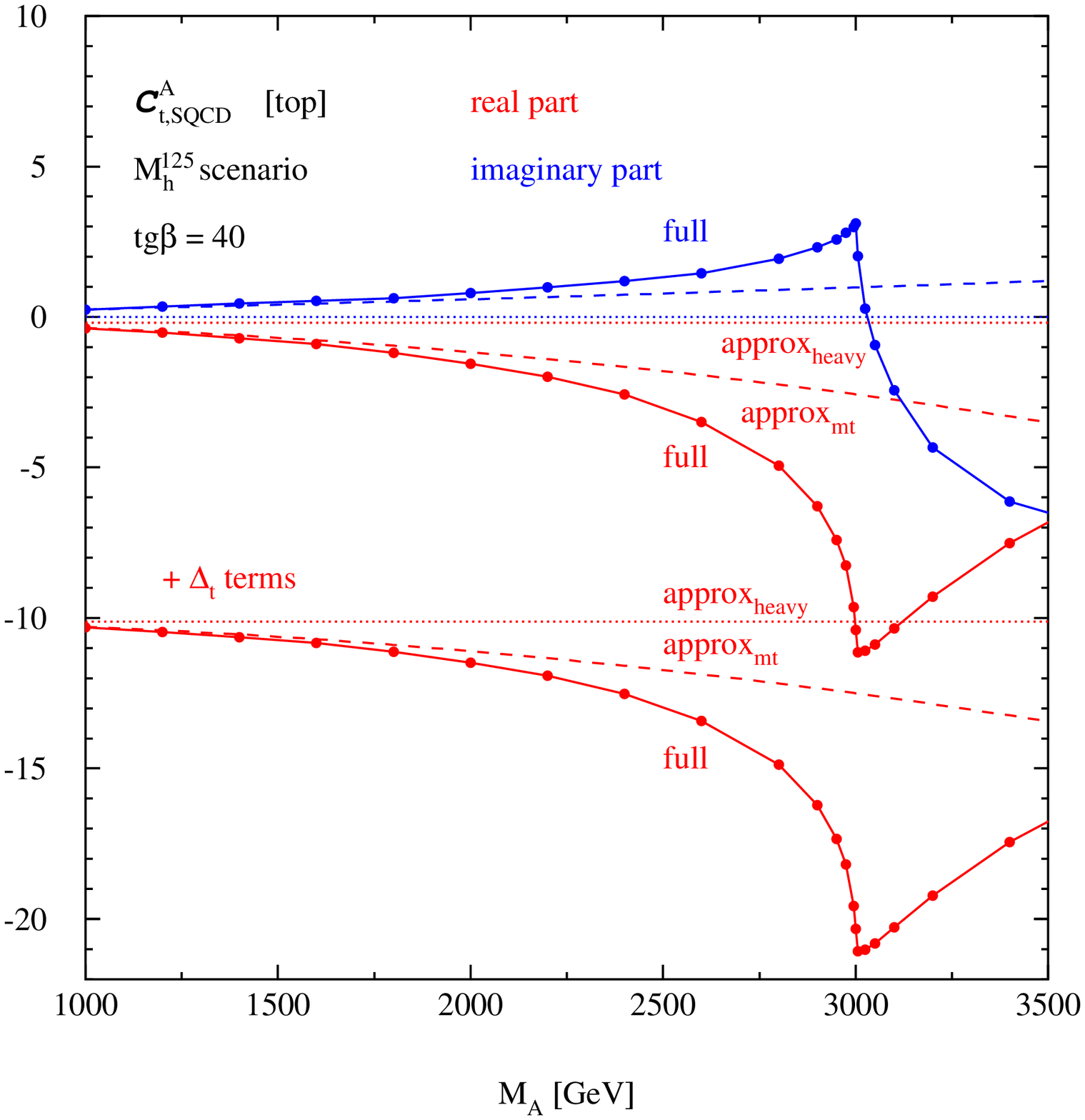}}
\put(60,-70.0){\includegraphics{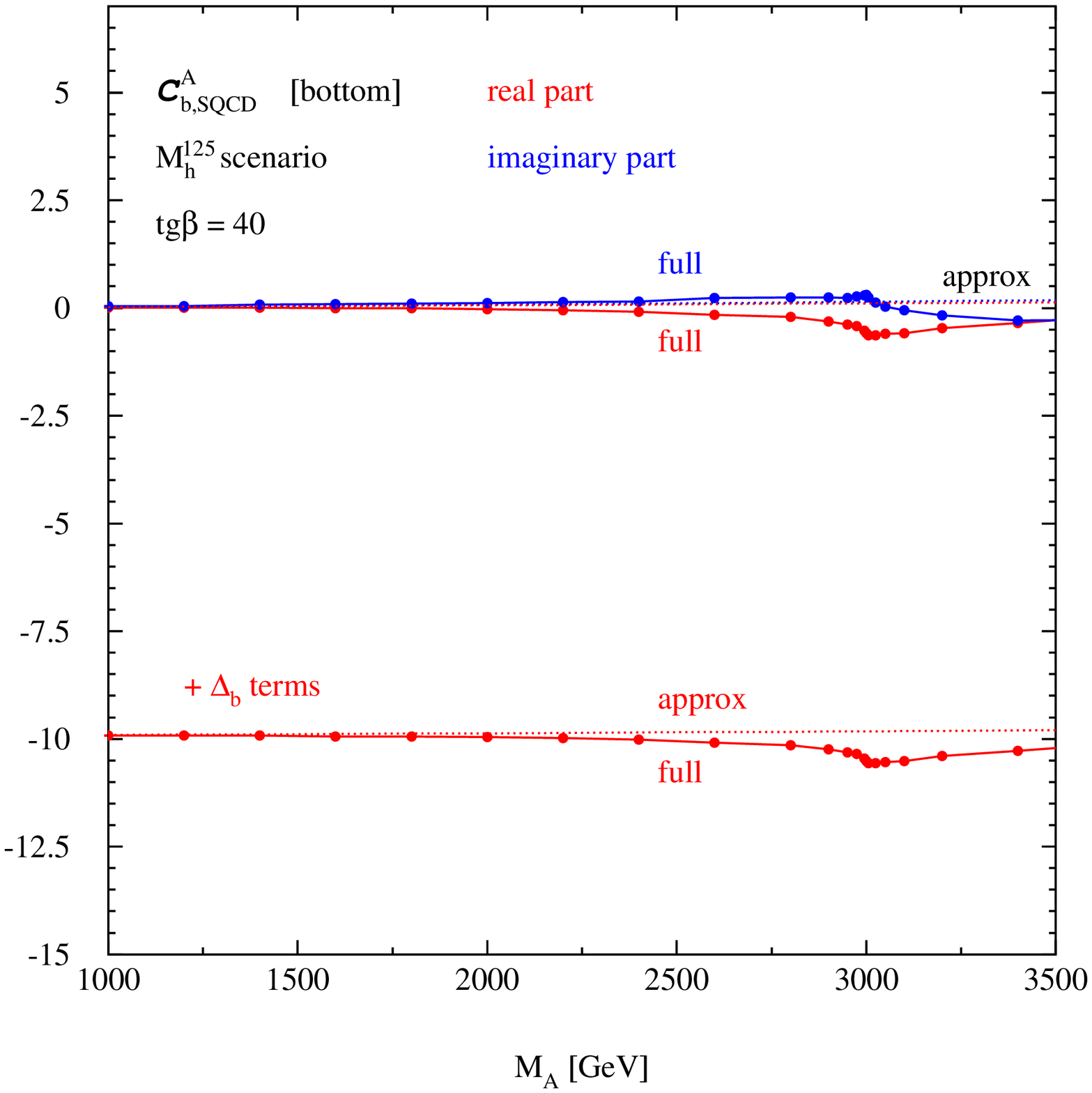}}
\end{picture}
\caption{\it The same as Fig.~\ref{fg:csusy10}, but for $\tgb=40$.}
\label{fg:csusy40}
\end{center}
\end{figure}
\begin{figure}[hbtp]
\begin{center}
\begin{picture}(150,230)(0,0)
\put(-180,-70.0){\includegraphics{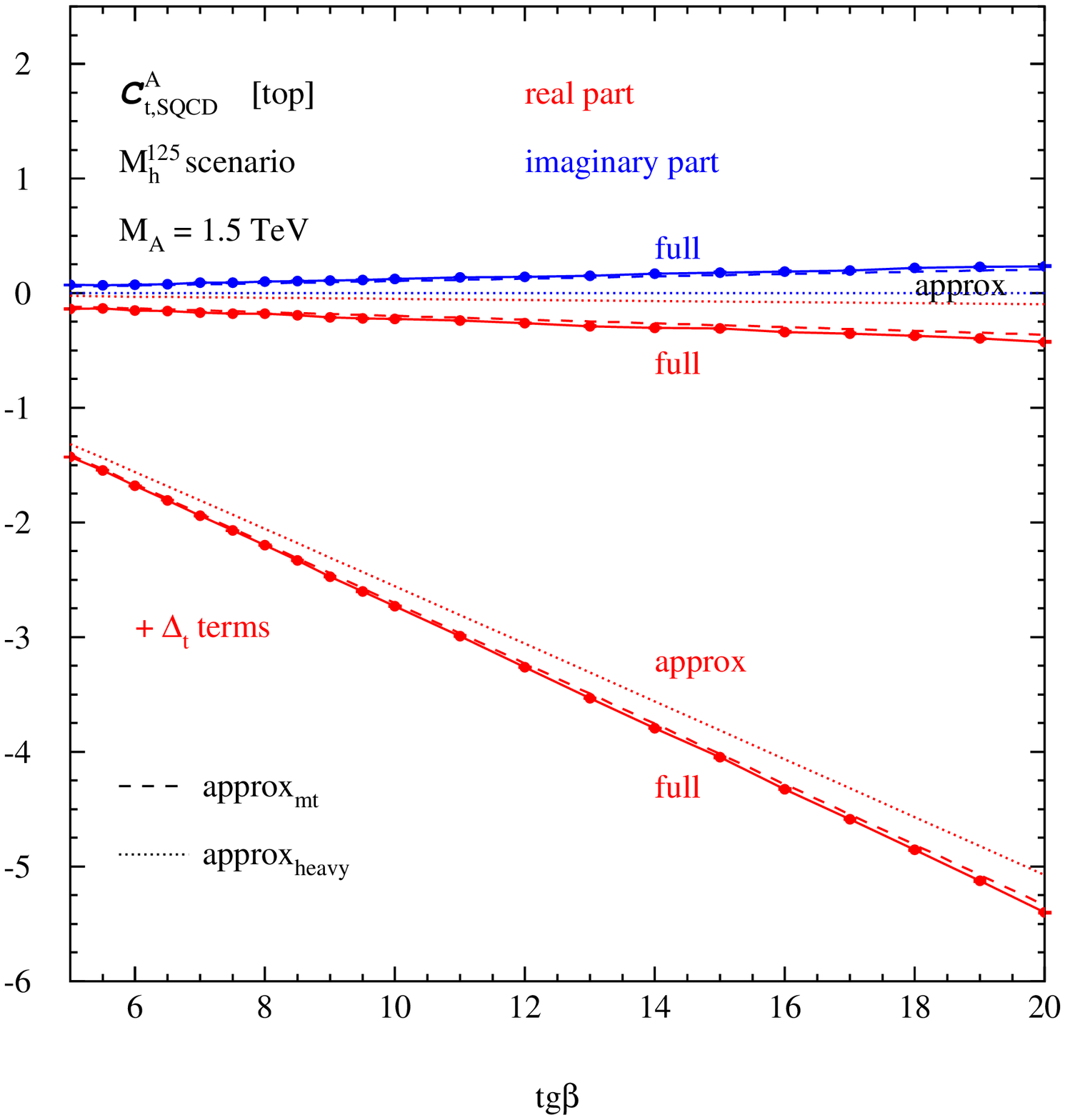}}
\put(60,-70.0){\includegraphics{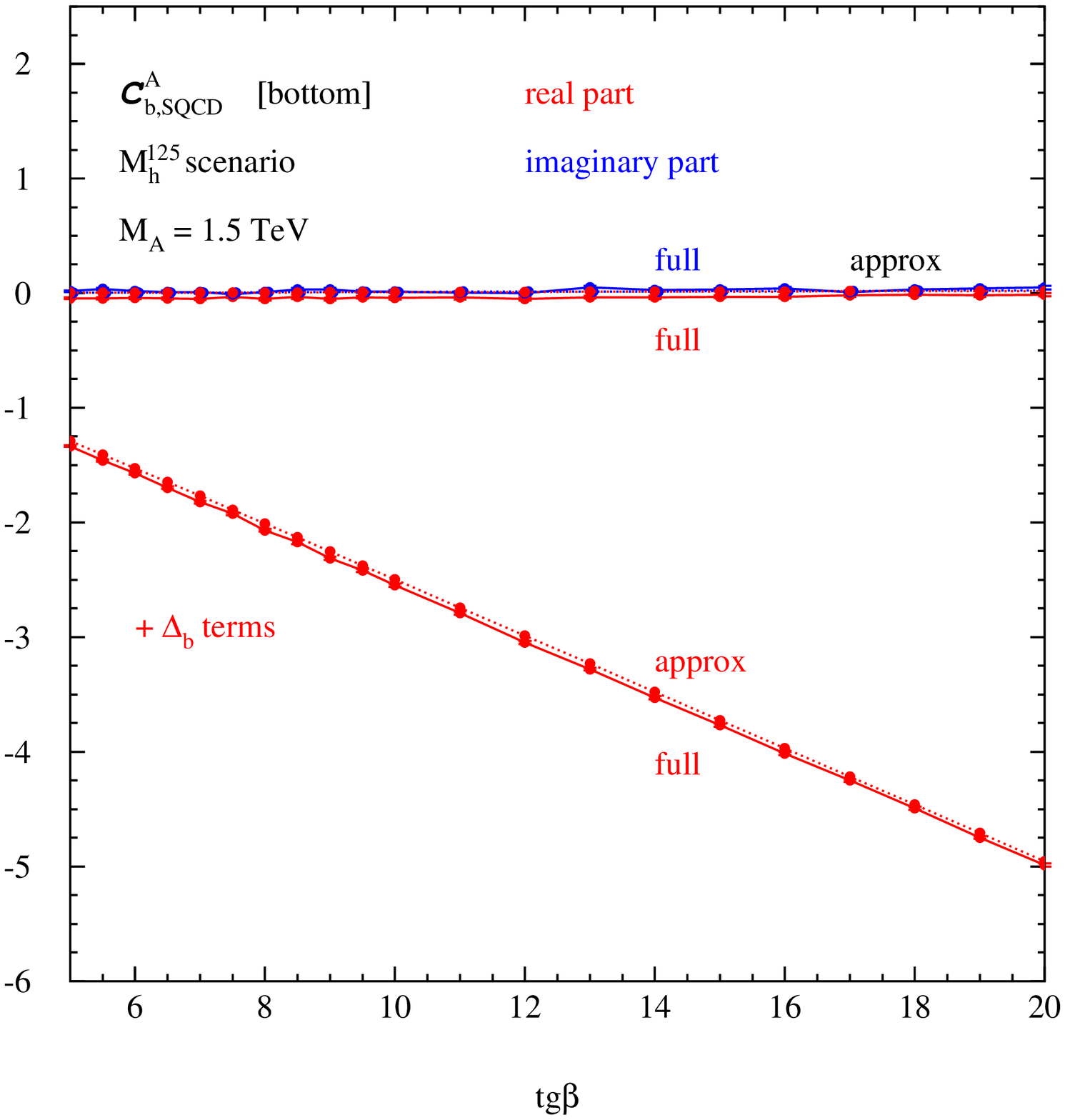}}
\end{picture}
\caption{\it The same as Fig.~\ref{fg:csusy10}, but as a function of
$\tgb$ for $M_A=1.5$ TeV.}
\label{fg:csusytgb}
\end{center}
\end{figure}
As a starting point, the perturbative NLO coefficients ${\cal C}^A_{Q,SQCD}$
are displayed in Figs.~\ref{fg:csusy10} and \ref{fg:csusy40} as a
function of the pseudoscalar mass $M_A$ for $\tgb$ values of 10 and 40,
respectively. In these figures, we show the approximate calculations of
Ref.~\cite{slavich} as well, i.e.~for the stop contribution both
approximations of the combined heavy-top/SUSY limit ('approx$_{heavy}$')
and the pure large SUSY-mass limit ('approx$_{mt}$'), while for the
sbottom contribution only the large SUSY-mass limit ('approx') is
phenomenologically relevant and shown. The full calculation agrees well
with the former approximate calculations for smaller pseudoscalar masses
in both the stop and sbottom cases. However, we observe sizeable and
increasingly relevant deviations for pseudoscalar masses approaching or
exceeding the virtual squark threshold\footnote{The kink structure at the heavy squark threshold is in line with the $S$-wave but ${\cal CP}$-odd behaviour of $\tilde q_1 \overline{\tilde q}_2$ and $\tilde q_2 \overline{\tilde q}_1$ pairs of different squarks close to the threshold.}. Moreover, we display the results of the NLO coefficients for the two cases of absorbing the
$\Delta_{t/b}$ terms in the corresponding Yukawa couplings and the
opposite. It is clearly visible that the $\Delta_{t/b}$ terms
approximate the full results quite well for smaller pseudoscalar masses
$M_A$ so that the results after subtracting them turn out to be quite
small. These subtracted results represent the SUSY-remainder, i.e.~the
contributions beyond the leading parts corresponding to the
effective top and bottom Yukawa couplings. It is obvious that the
absorption of these contributions leads to a much better perturbative
behaviour thus corroborating the effective Yukawa-coupling approach.
This is further underlined by the $\tgb$ dependence of the stop and
sbottom contributions shown in Fig.~\ref{fg:csusytgb} for a pseudoscalar
mass $M_A=1.5$ TeV. The description
of the SUSY--QCD corrected cross section in terms of the effective
low-energy top- and bottom-Yukawa couplings leads to a moderate
SUSY-remainder at NLO as long as the pseudoscalar Higgs mass does not
approach the virtual stop/sbottom thresholds. At and beyond these virtual
thresholds, the SUSY-remainders turn out to be sizeable.

\begin{figure}[hbtp]
\begin{center}
\begin{picture}(150,230)(0,0)
\put(-180,-70.0){\includegraphics{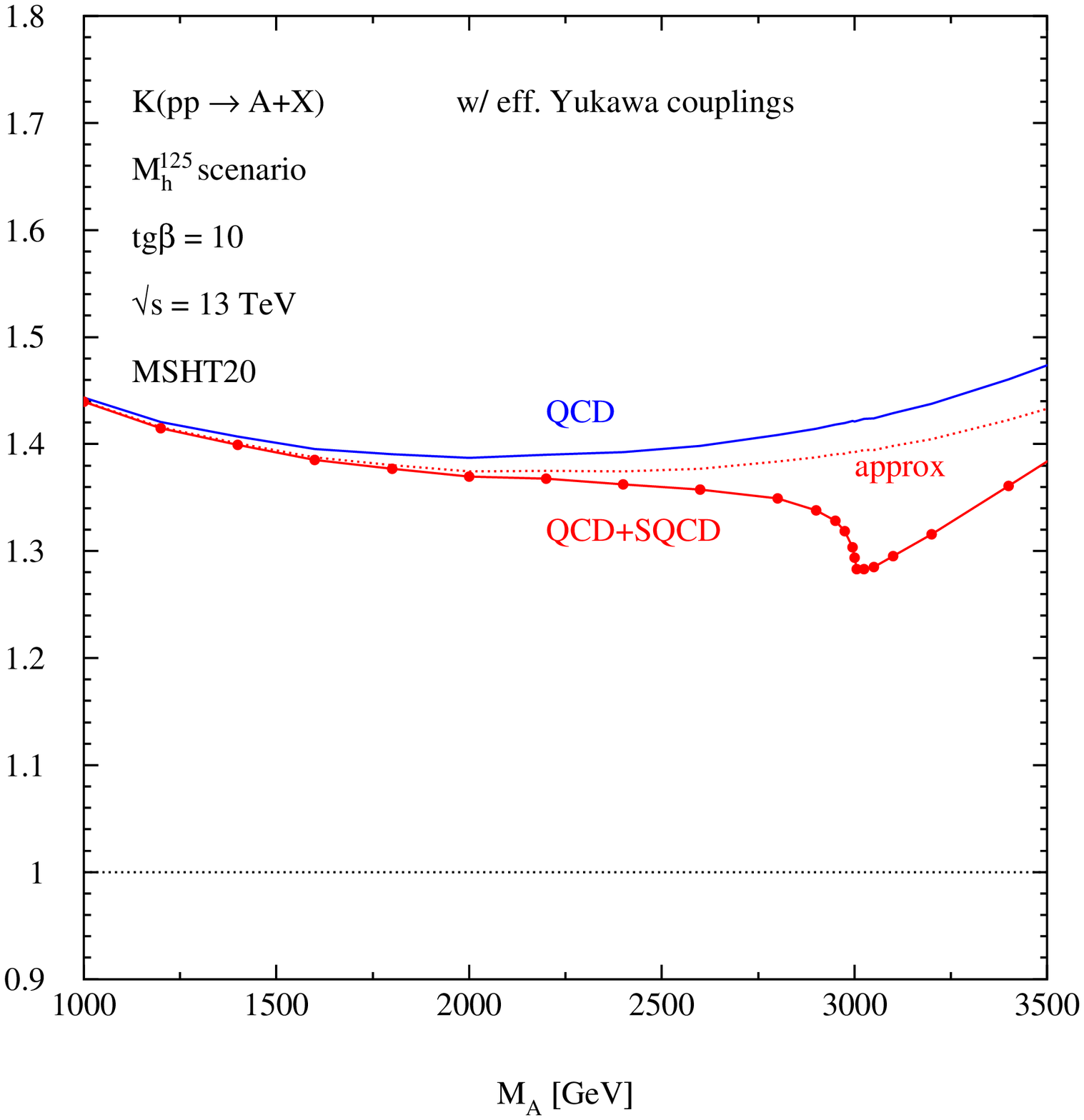}}
\put(60,-70.0){\includegraphics{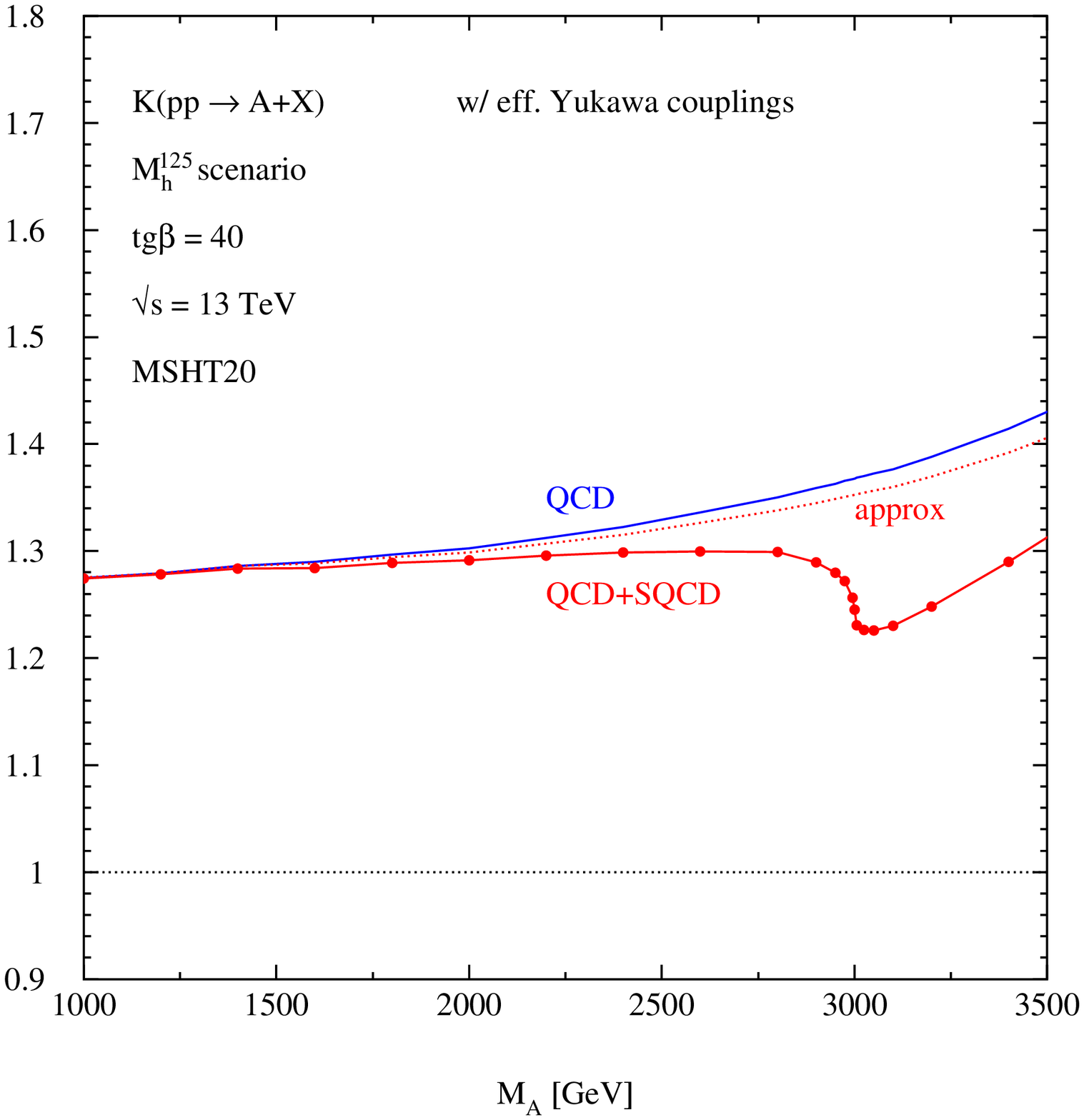}}
\end{picture}
\caption{\it The K-factors of the QCD and genuine SUSY--QCD corrections
for the LHC with $\tgb=10, 40$ and a c.m.~energy of 13 TeV. As parton
density functions the MSHT20 sets have been used. The renormalization
and factorization scales have been chosen as $\mu_R = \mu_F = M_A/2$.}
\label{fg:kfac}
\end{center}
\end{figure}
\begin{figure}[hbt]
\begin{center}
\begin{picture}(150,230)(0,0)
\put(-180,-70.0){\includegraphics{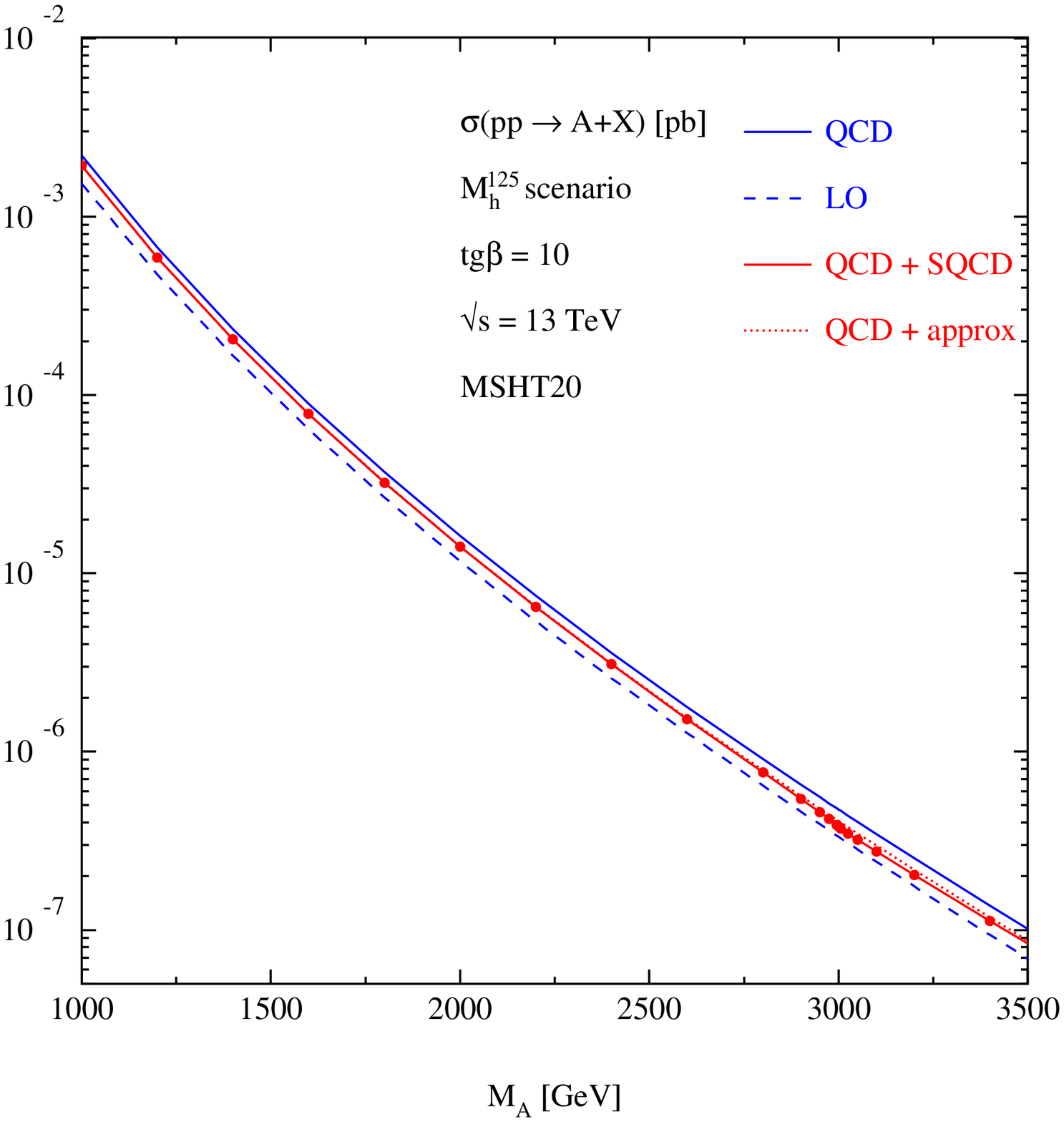}}
\put(60,-70.0){\includegraphics{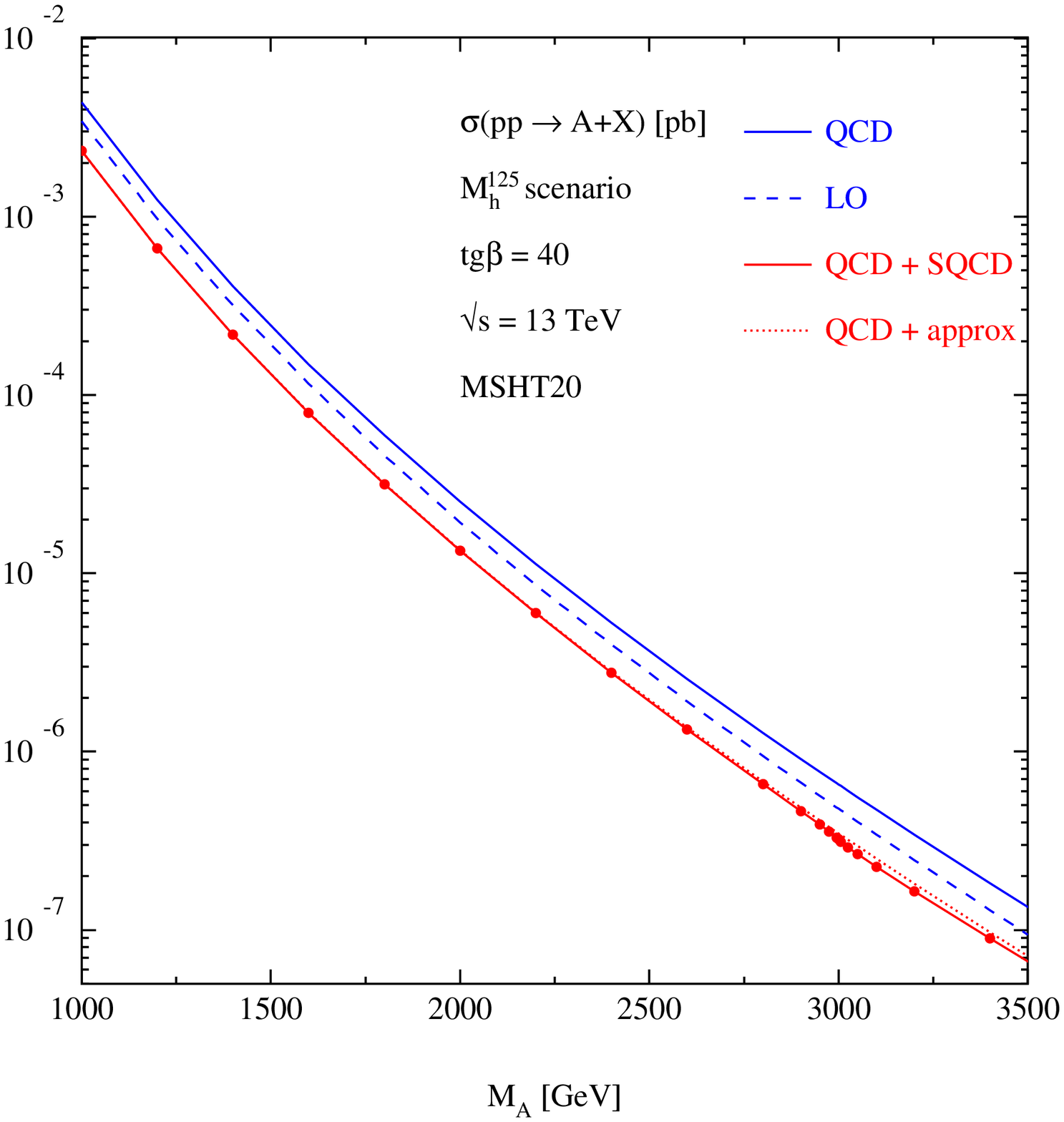}}
\end{picture}
\caption{\it The pseudoscalar production cross section via gluon-fusion
at the LHC with $\tgb=10$ (left) and $\tgb=40$ (right) and a
c.m.~energy of 13 TeV at LO, NLO QCD and including the genuine SUSY--QCD
corrections involving effective Yukawa couplings. The LO and NLO QCD
corrected cross sections are shown without effective Yukawa couplings. As
parton density functions the MSHT20 sets have been used. The
renormalization and factorization scales have been chosen as $\mu_R =
\mu_F = M_A/2$.}
\label{fg:cxn}
\end{center}
\end{figure}
As the next step, we analyze the SUSY--QCD corrections to the hadronic
cross section of pseudoscalar Higgs-boson production via gluon fusion.
The effect of the corrections on the $K$-factor at the hadronic level,
which is defined as the ratio between the NLO and LO cross sections, is
discussed first. We adopt the {\tt MSHT20nlo\_as118} parton density functions and
perform the analysis for a c.m.~energy of 13 TeV at the LHC.
Fig.~\ref{fg:kfac} exhibits the $K$-factor for $\tgb=10,40$ with effective
top- and bottom-Yukawa couplings for the QCD part of the cross section
and for the corresponding results of the previous approximate
calculations. The QCD part of the $K$-factors shows the usual sizeable NLO corrections of about 30--50\%, while the additional SUSY--QCD remainder turns out to be small or moderate. The comparison implies that effects beyond the approximation become relevant when approaching the virtual stop/sbottom
thresholds and above as expected.

These $K$-factors can be translated to the hadronic production cross
sections of pseudoscalar Higgs bosons via gluon fusion as shown in
Fig.~\ref{fg:cxn} for two values of $\tgb=10,40$. For the effective
bottom-Yukawa couplings, we include the full set of NNLO corrections
\cite{deltabnlo} to lift the accuracy of the factorizing and dominant
contributions to the NNLO level, while for the effective top-Yukawa
coupling we use the NLO expression in the effective field-theory
framework. Here, we present the QCD-corrected cross sections without the
effective top- and bottom-Yukawa couplings, i.e.~without any genuine SUSY--QCD corrections and the approximate and full
SUSY--QCD corrected cross sections with the effective Yukawa couplings
as discussed in the previous section. The comparison of the full
QCD-corrected cross section (blue line) and the full QCD + SUSY--QCD
corrected cross section (red line) supports the high relevance of the
SUSY--QCD corrections in total, while the SUSY-remainder plays a role
close or above the virtual stop- and sbottom thresholds.

\subsection{The Gluonic Decay $A\to gg$}
\begin{figure}[hbtp]
\begin{center}
\begin{picture}(150,225)(0,0)
\put(-180,-70.0){\includegraphics{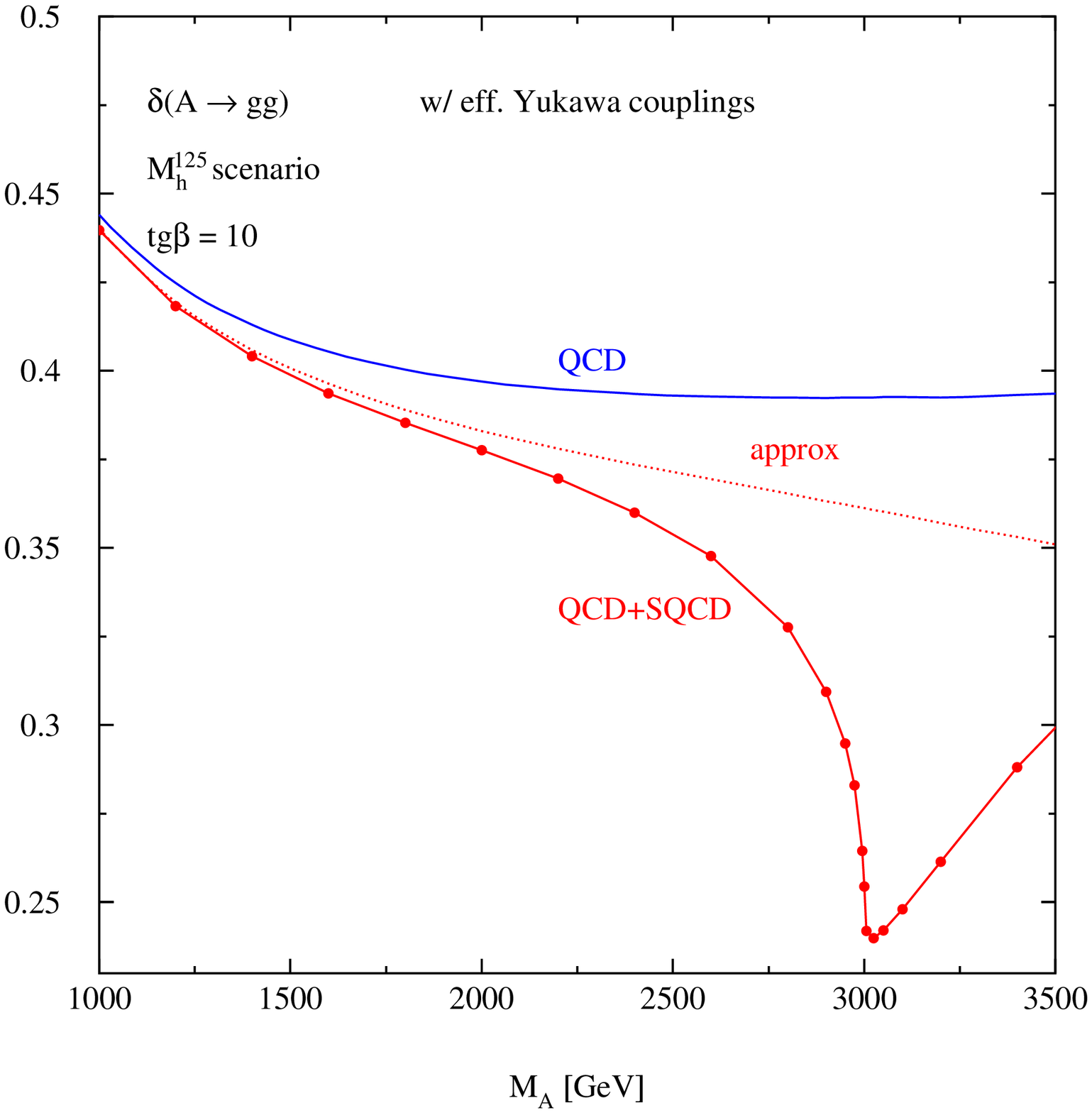}}
\put(60,-70.0){\includegraphics{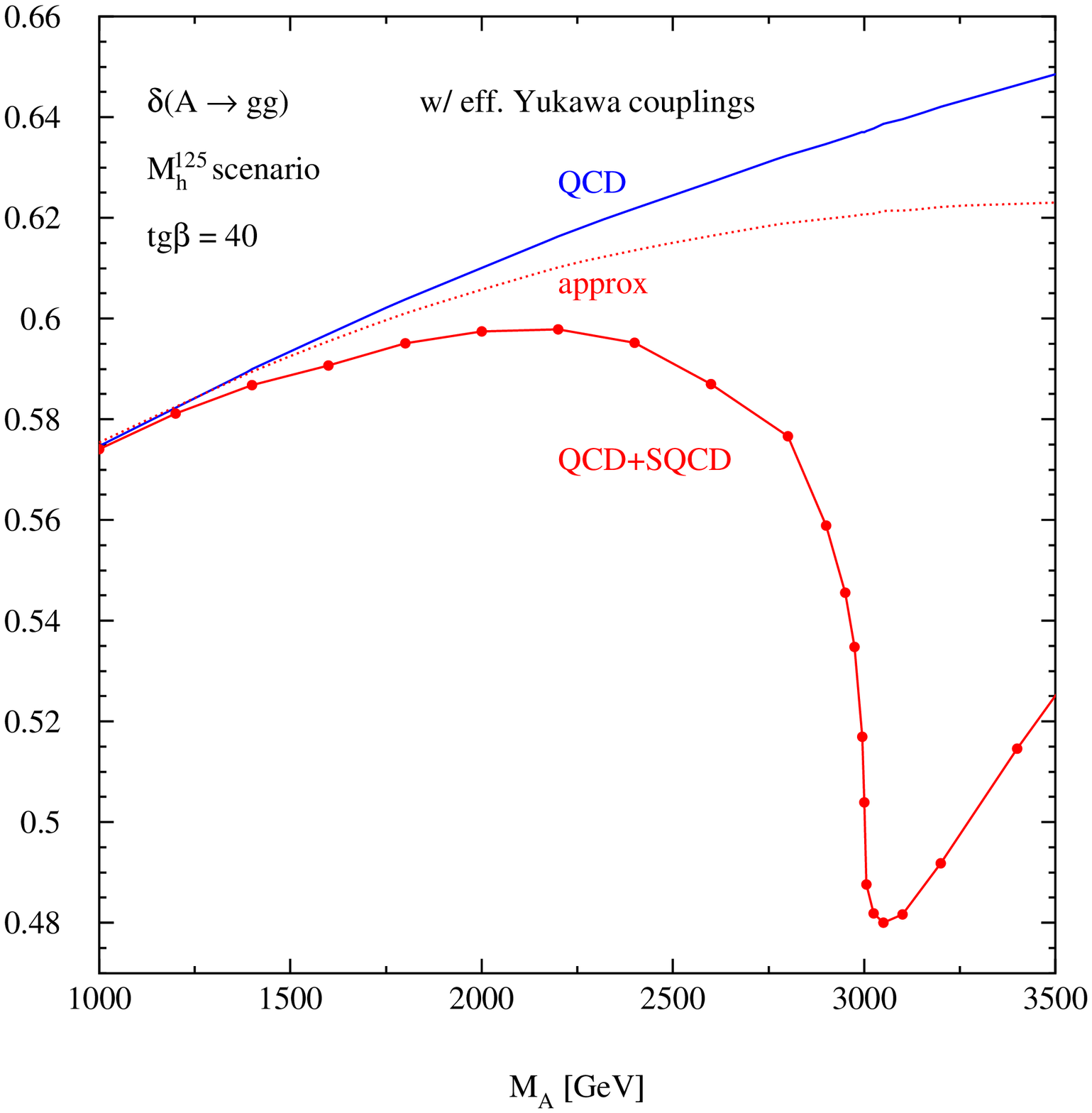}}
\end{picture}
\caption{\it Relative radiative corrections, defined as $\Gamma =
\Gamma_{LO} (1+\delta)$, to the gluonic pseudoscalar decay width as a
function of the pseudoscalar mass $M_A$ for $\tgb=10$ (left) and
$\tgb=40$ (right) at NLO QCD and including the genuine SUSY--QCD
corrections involving effective Yukawa couplings. The renormalization
scale has been chosen as $\mu_R = M_A$.}
\label{fg:a2gg_corr}
\end{center}
\end{figure}

The same virtual coefficient as for $gg\to A$ contributes to the genuine
SUSY--QCD corrections of the gluonic pseudoscalar Higgs decay $A\to gg$
according to Eq.~(\ref{eq:a2ggsqcd}). The relative QCD and SUSY--QCD
corrections to the gluonic decay width are shown in
Fig.~\ref{fg:a2gg_corr} with the use of effective top and
bottom Yukawa couplings. It is clearly
visible that the bulk of the 
genuine SUSY--QCD corrections can be absorbed by the effective top and
bottom Yukawa couplings including $\Delta_{t,b}$ contributions. The
SUSY-remainder is relevant in regions where finite squark-mass effects
become relevant, i.e.~close or above the related virtual thresholds.
\begin{figure}[hbtp]
\begin{center}
\begin{picture}(150,225)(0,0)
\put(-180,-70.0){\includegraphics{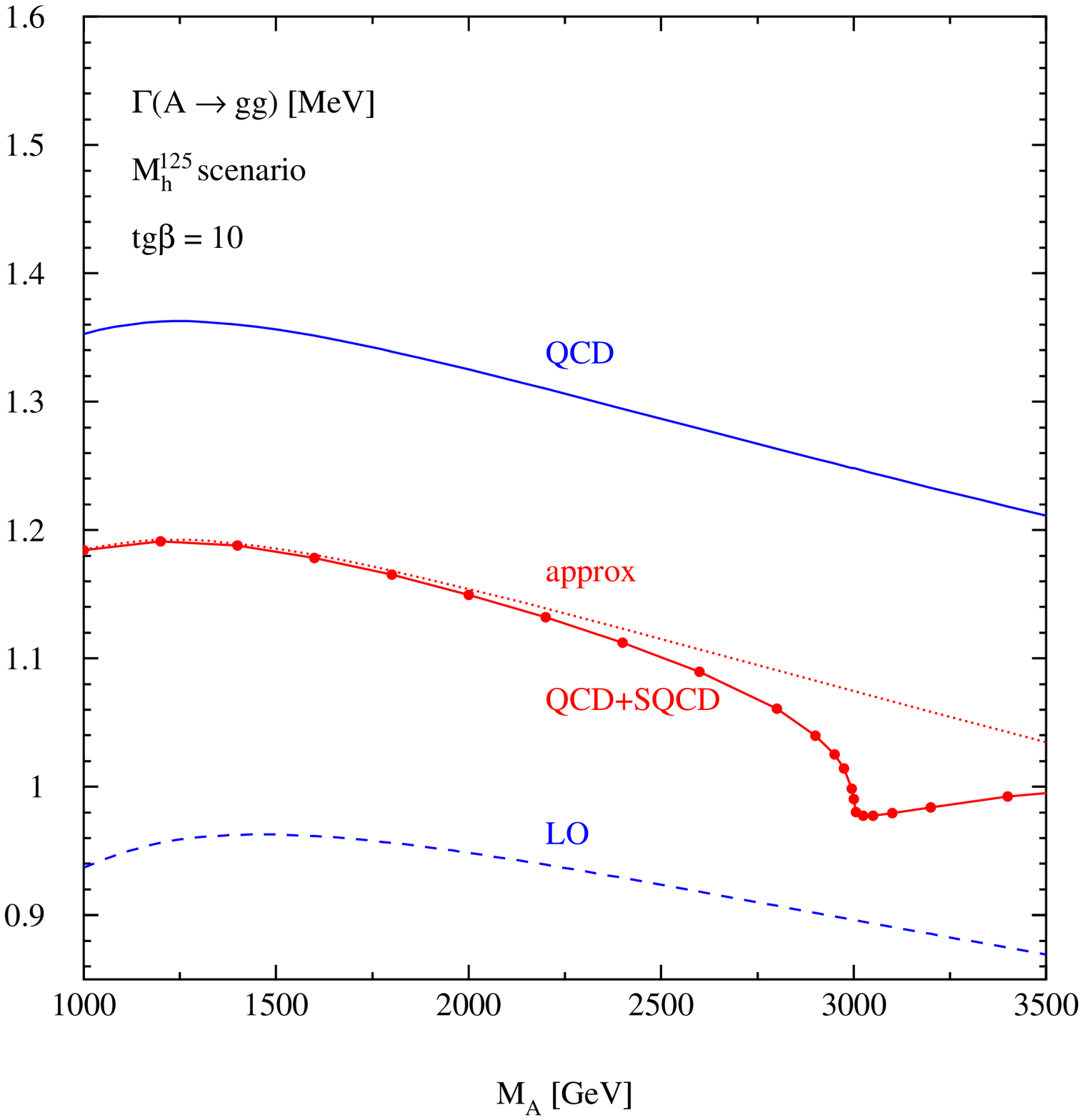}}
\put(60,-70.0){\includegraphics{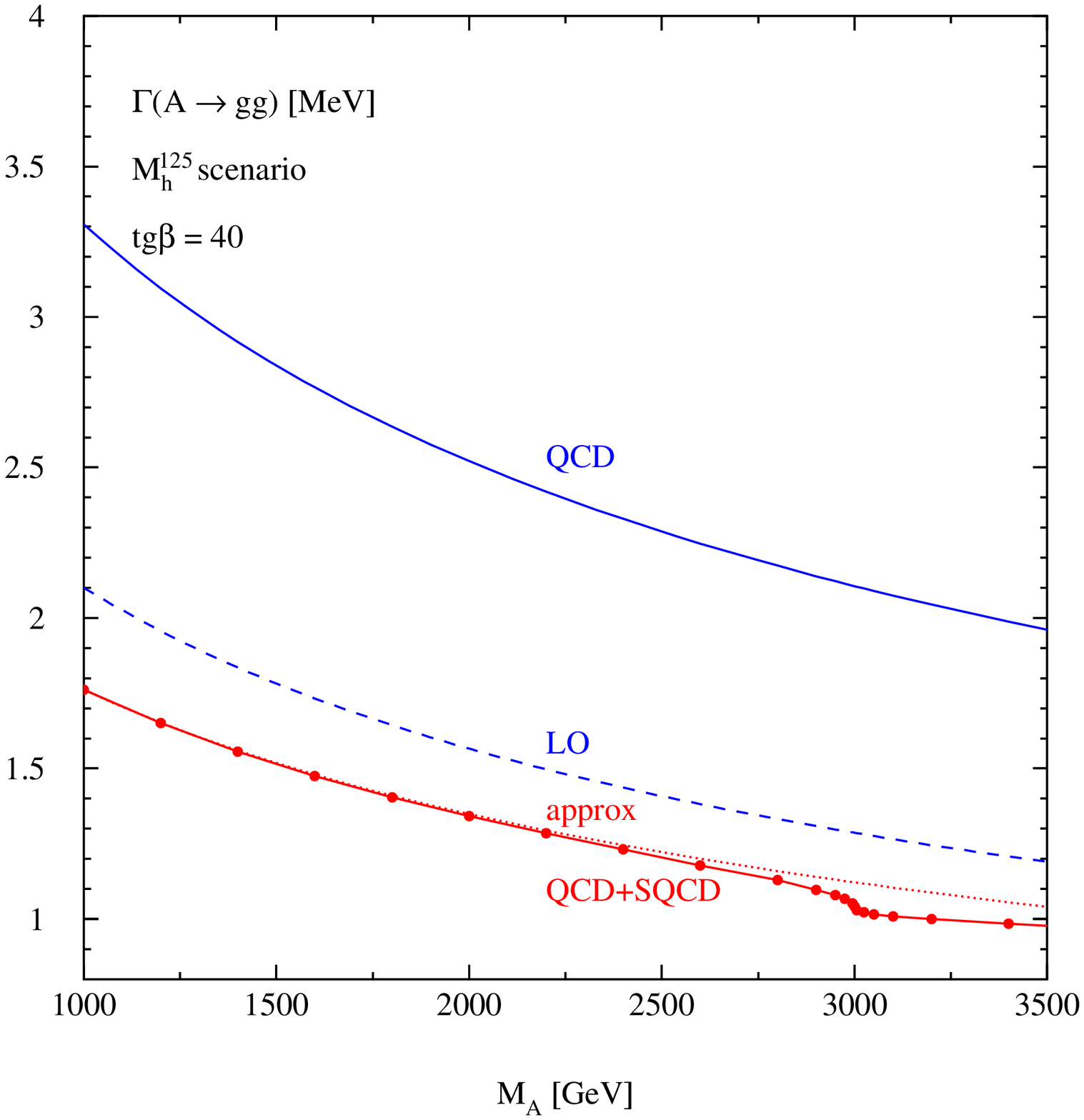}}
\end{picture}
\caption{\it Partial decay widths $\Gamma(A\to gg)$ as a
function of the pseudoscalar mass $M_A$ for $\tgb=10$ (left) and
$\tgb=40$ (right) at NLO QCD and including the genuine SUSY--QCD
corrections involving effective Yukawa couplings. The renormalization
scale has been chosen as $\mu_R = M_A$.}
\label{fg:a2gg}
\end{center}
\end{figure}
The corresponding partial decay widths $\Gamma(A\to gg)$ are shown in Fig.~\ref{fg:a2gg} for
$\tgb=10,40$, using effective top and bottom Yukawa couplings for the SUSY--QCD-corrected decay
widths, but LO couplings without $\Delta_{t.b}$ terms for the LO and
QCD-corrected decay widths.
The SUSY--QCD corrections are treated in the same way as for the production cross sections,
i.e.~$\Delta_b$ terms at two-loop order and $\Delta_t$ contributions at one-loop level. The
main effect of the genuine SUSY--QCD corrections emerges from the factorizing $\Delta_{b,t}$
corrections to the Yukawa couplings. The comparison of the pure NLO QCD prediction (blue curve) and the SUSY--QCD corrected one (red curve) indicates the large size of SUSY--QCD corrections at NLO for the partial width.

\subsection{The Photonic Decay $A\to \gamma\gamma$} \label{sc:a2gaga}
\begin{figure}[hbtp]
\begin{center}
\begin{picture}(150,240)(0,0)
\put(-180,-70.0){\includegraphics{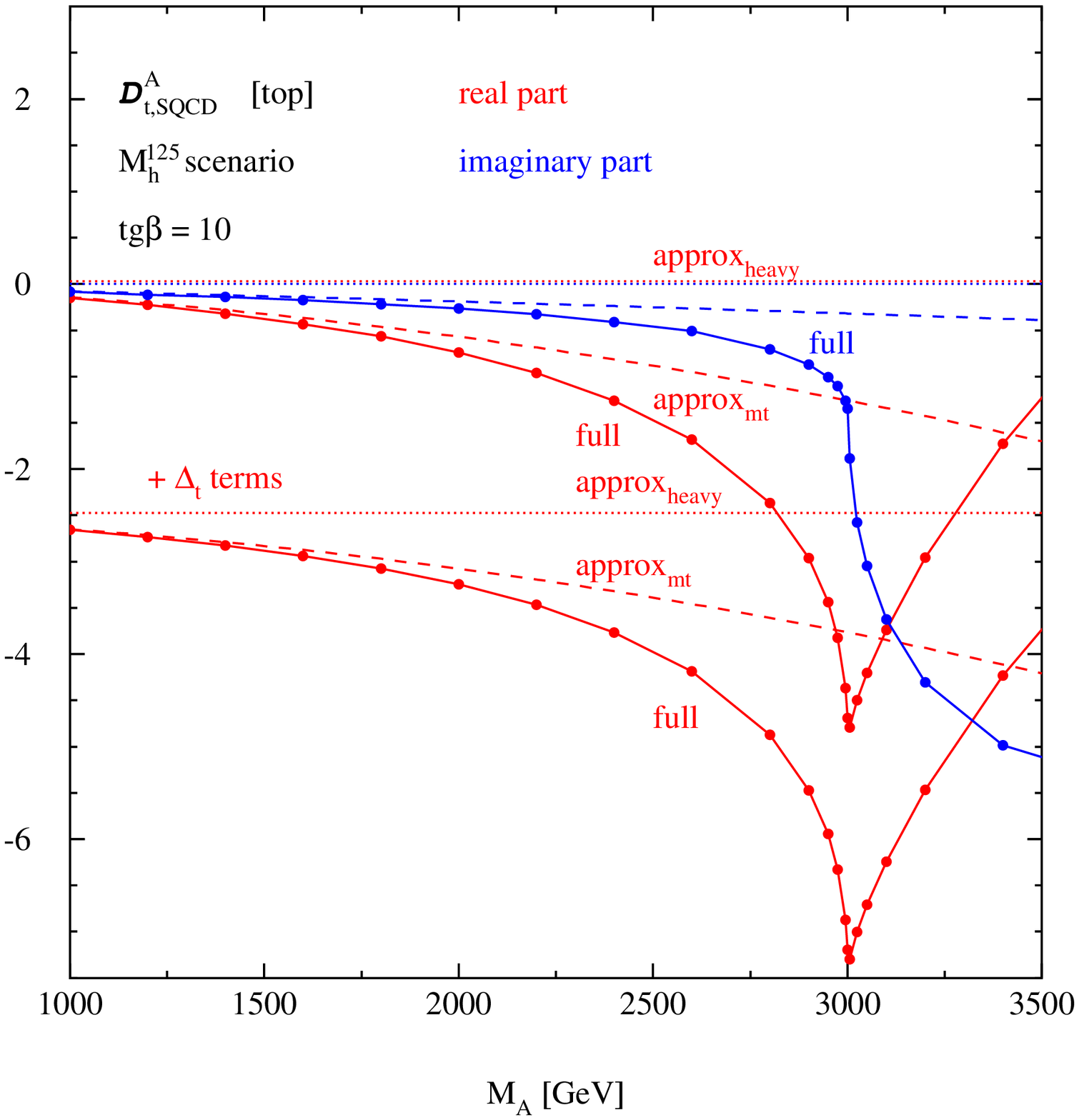}}
\put(60,-70.0){\includegraphics{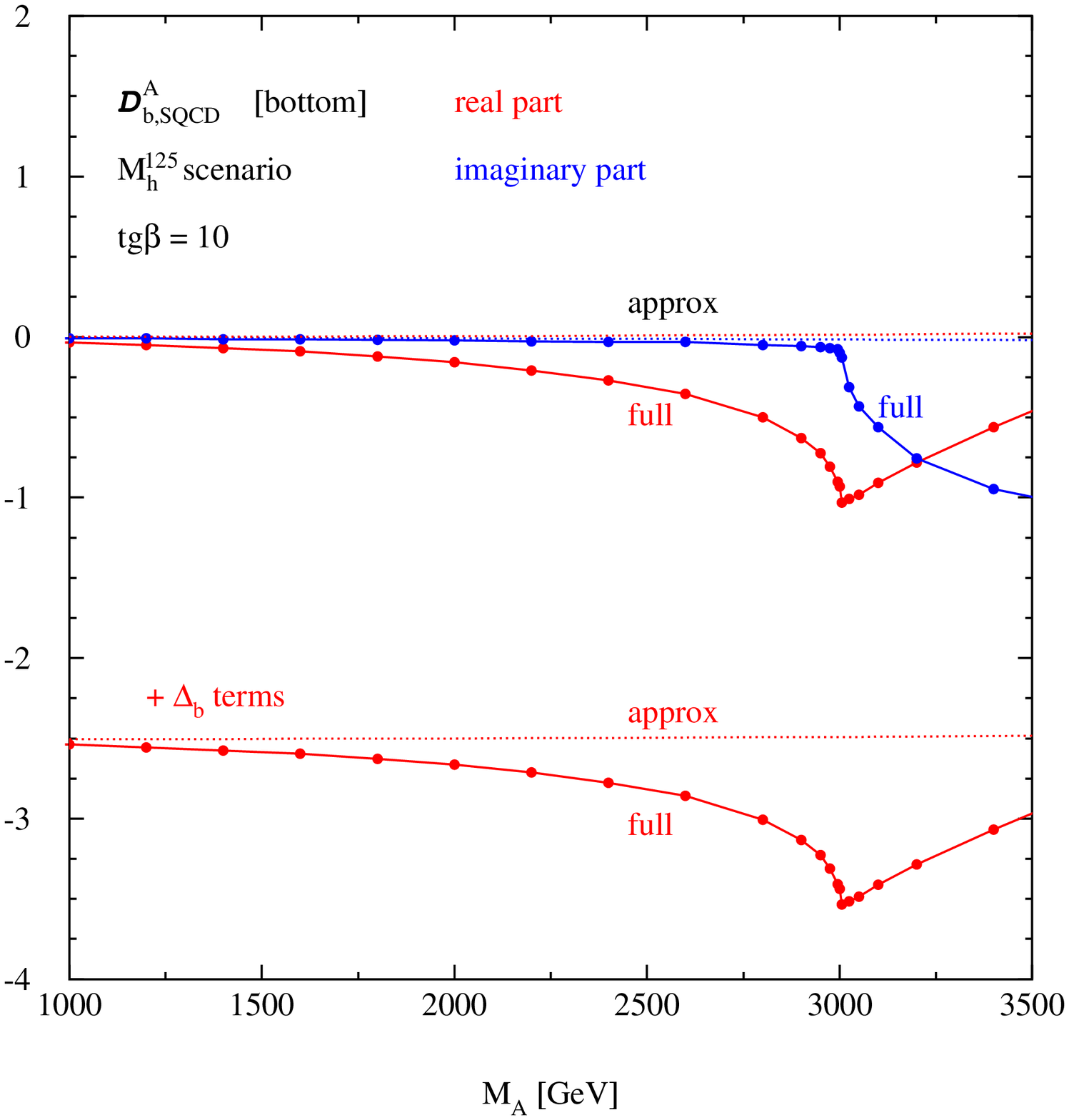}}
\end{picture}
\caption{\it The genuine SUSY--QCD corrections to $A\to\gamma\gamma$
normalized to the LO top and bottom quark form factors for $\tgb=10$ in
the $M_h^{125}$ benchmark scenario. Real part: red, imaginary part:
blue, compared to the Abelian part of the approximate calculations of
Ref.~\cite{slavich} (dashed lines). The dotted lines for the stop
contributions correspond the combined limit of large top and SUSY
masses.}
\label{fg:dsusy10}
\end{center}
\end{figure}

\begin{figure}[hbtp]
\begin{center}
\begin{picture}(150,230)(0,0)
\put(-180,-70.0){\includegraphics{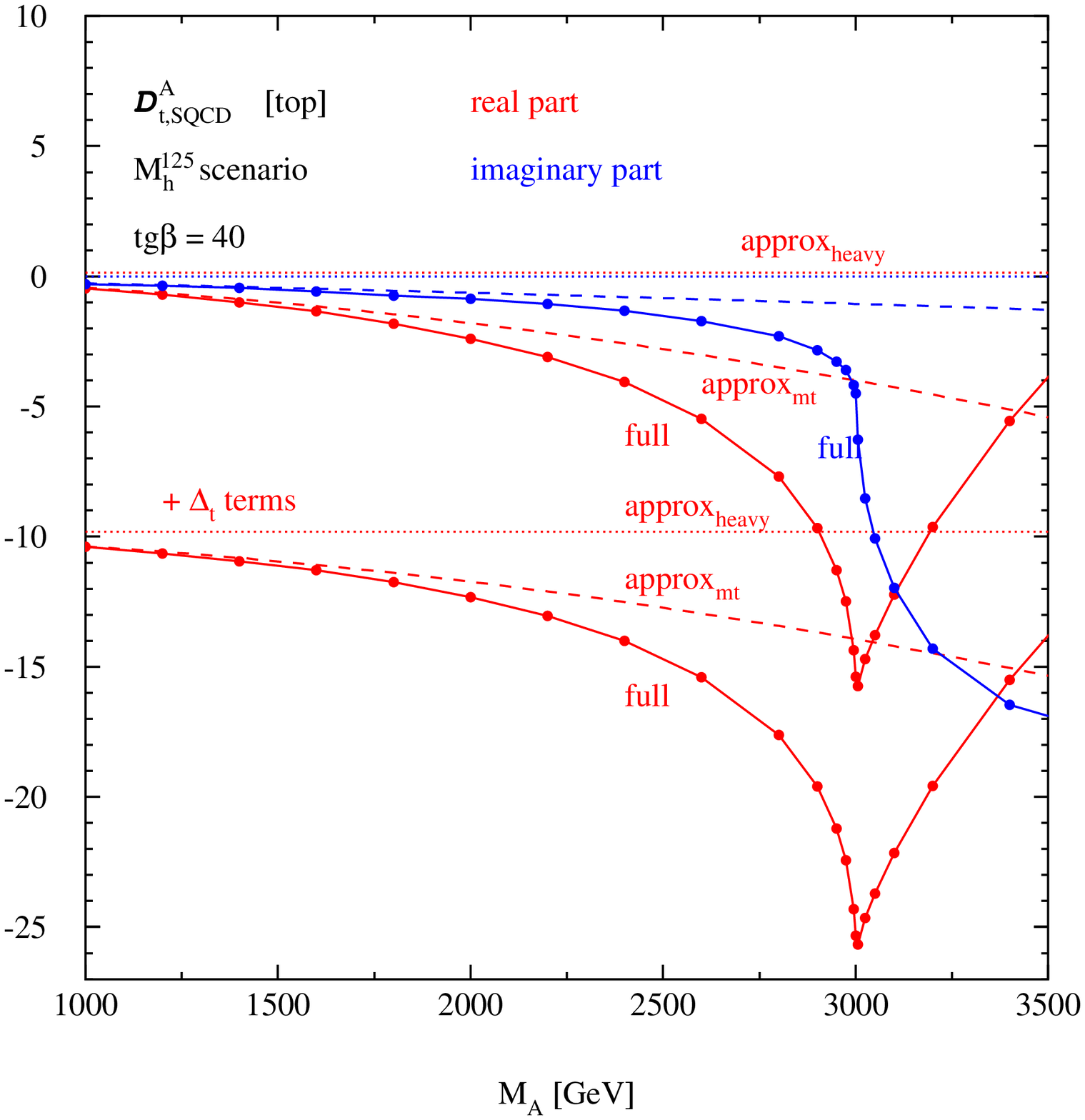}}
\put(60,-70.0){\includegraphics{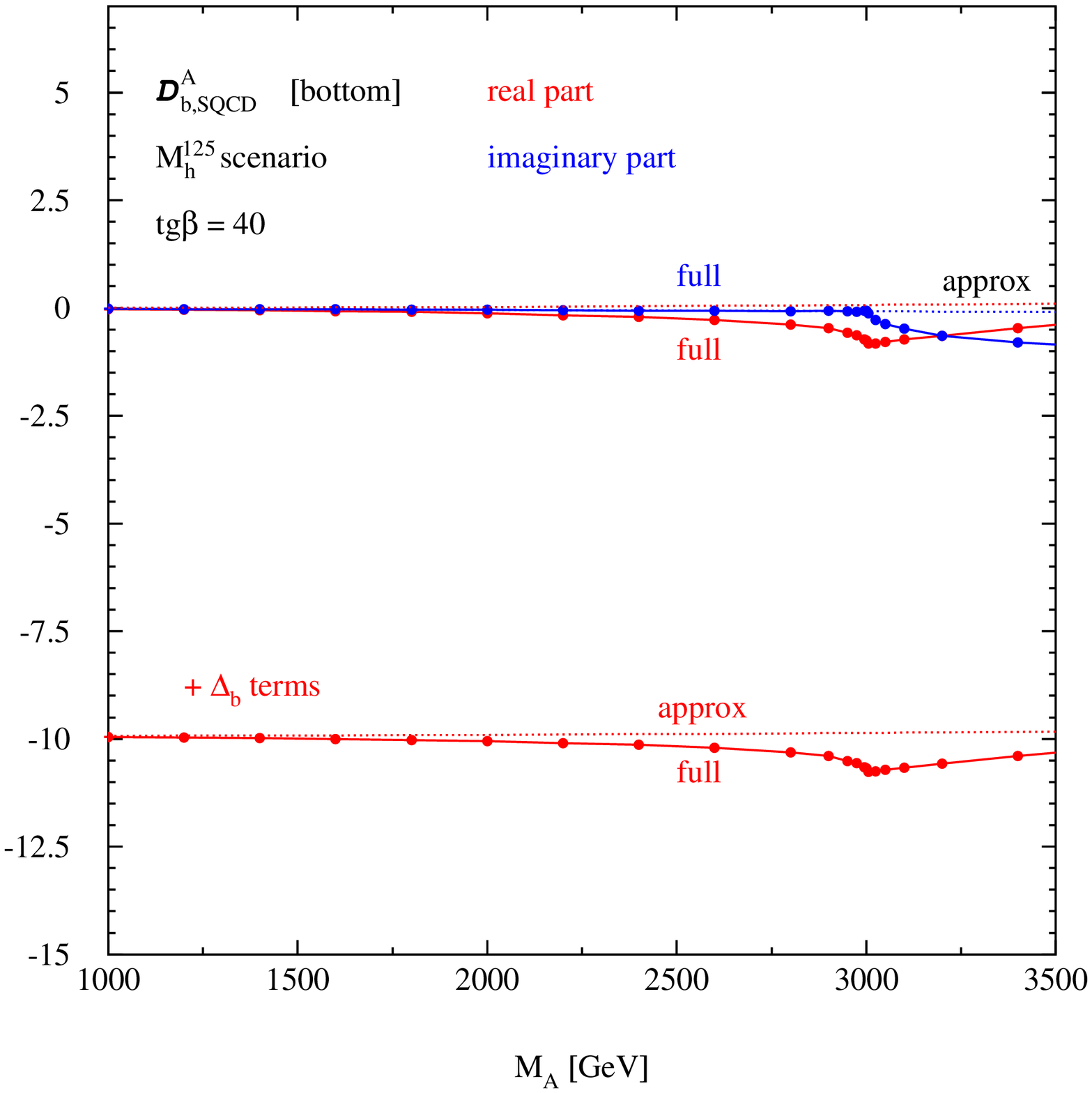}}
\end{picture}
\caption{\it The same as Fig.~\ref{fg:dsusy10}, but for $\tgb=40$.}
\label{fg:dsusy40}
\end{center}
\end{figure}

\begin{figure}[hbtp]
\begin{center}
\begin{picture}(150,220)(0,0)
\put(-180,-70.0){\includegraphics{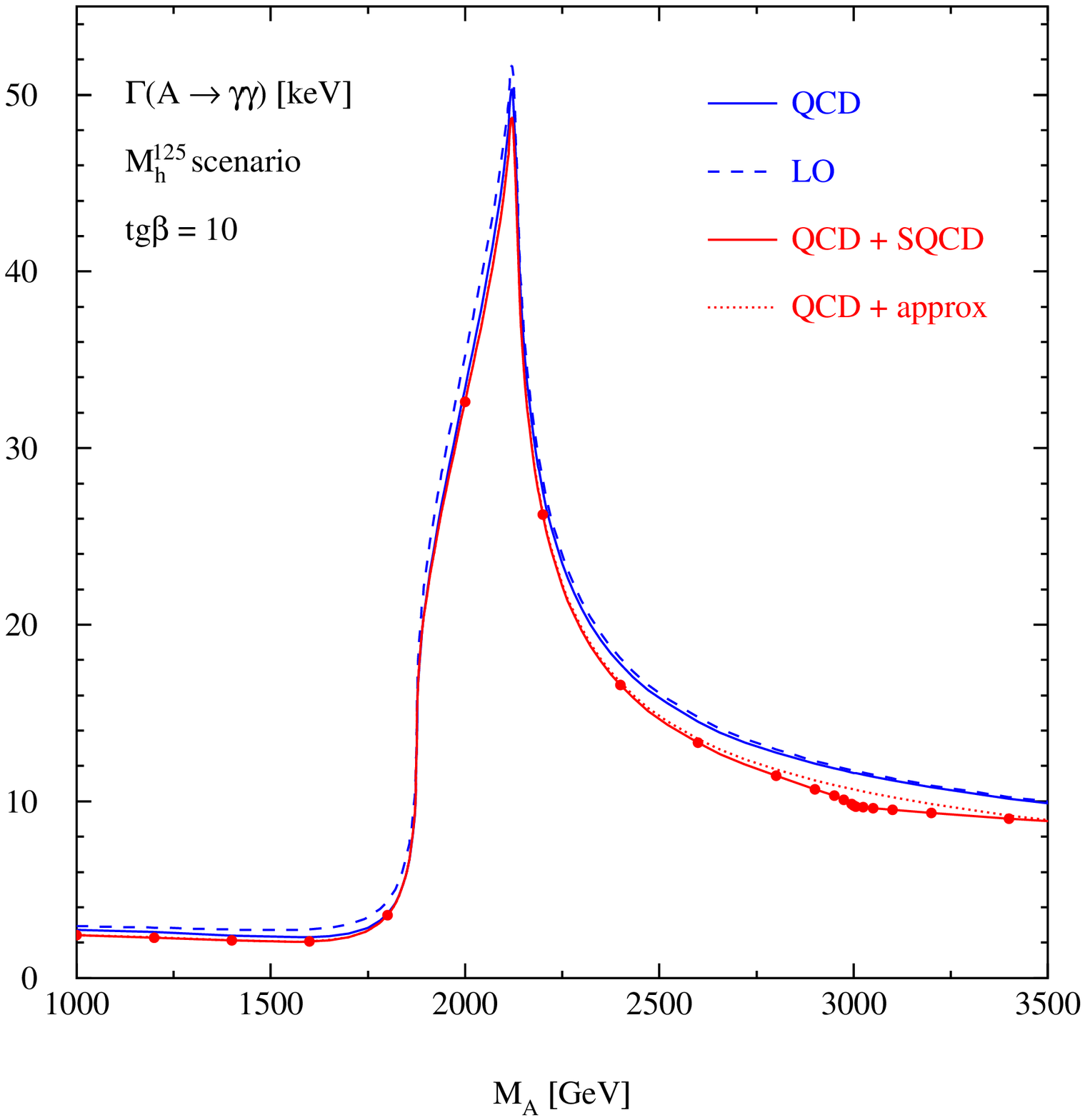}}
\put(60,-70.0){\includegraphics{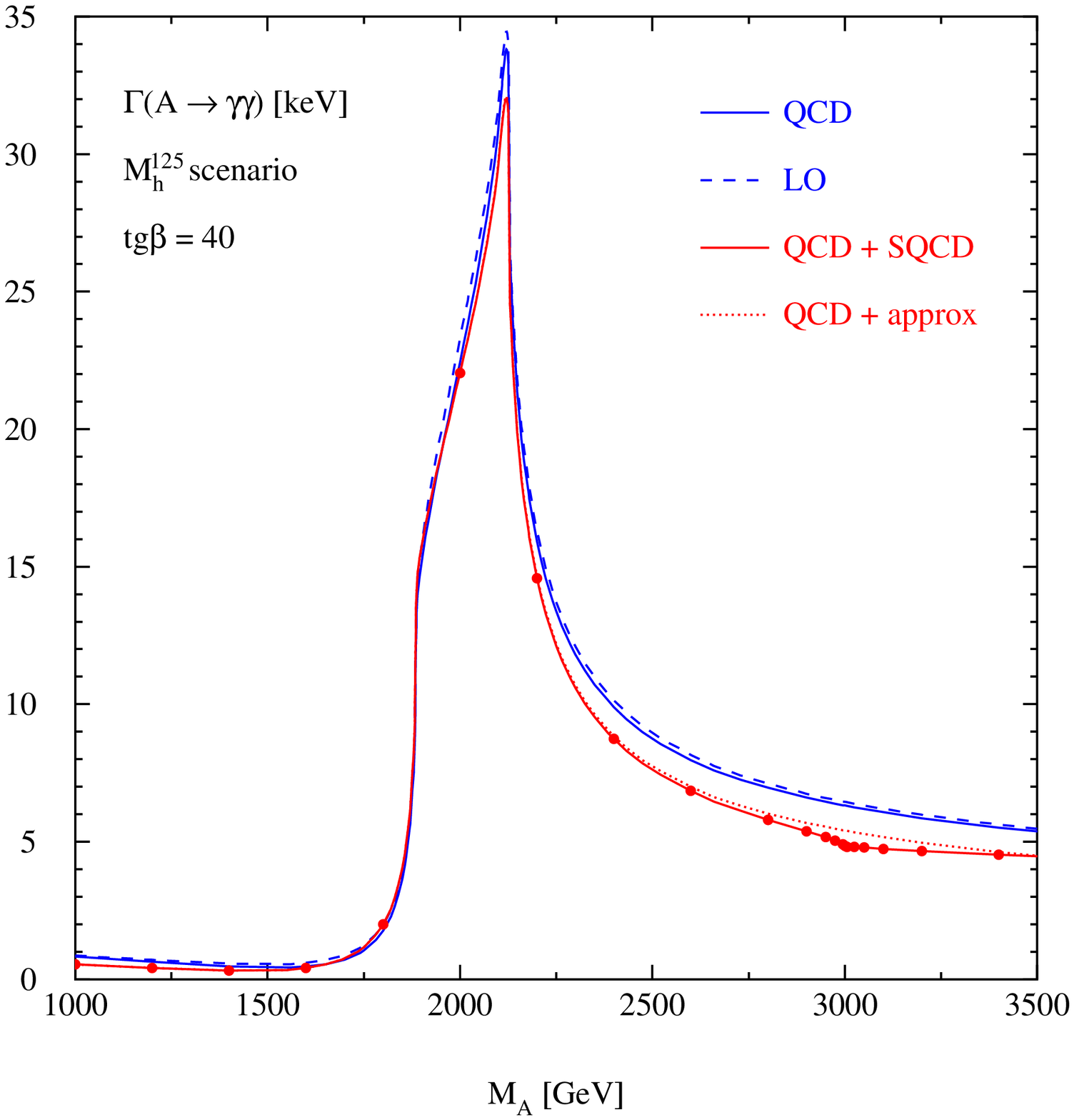}}
\end{picture}
\caption{\it The partial decay width of $A\to\gamma\gamma$ for two
  values of $\tgb=10,40$ in the $M_h^{125}$ benchmark scenario at LO
  and NLO QCD (blue) and including the genuine SUSY--QCD corrections
  involving effective Yukawa couplings (red).}
\label{fg:gam_gaga}
\end{center}
\end{figure}

The virtual SUSY--QCD corrections to the photonic decay width of
$A\to\gamma\gamma$ emerge from the first four diagrams of
Fig.~\ref{fg:nlodia} after adjusting the related coupling and color
factors and replacing the two external gluons by photons. The normalized
coefficient of the SUSY--QCD corrections with and without absorption of
the $\Delta_{t,b}$ terms is shown in Figs.~\ref{fg:dsusy10} and
\ref{fg:dsusy40} for two values of $\tgb = 10,40$. As in the gluonic case the $\Delta_t$ and $\Delta_b$ contributions determine the dominant part of the SUSY--QCD corrections that can be absorbed in the effective top and bottom Yukawa couplings of Eq.~(\ref{eq:rescoup}). The SUSY--QCD remainder turns out to be small apart from the regions closer to the virtual stop and sbottom thresholds. The partial decay widths of $A\to\gamma\gamma$ are shown in Fig.~\ref{fg:gam_gaga} for the different levels of perturbative orders. The LO and NLO QCD corrected widths are shown in blue, while the approximate and full SUSY--QCD-corrected ones are displayed in red. As in the previous cases it is clearly visible that the bulk of the genuine
SUSY--QCD corrections can be absorbed by the corresponding effective top
and bottom Yukawa couplings leaving a sizeable SUSY-remainder in regions
only where squark-mass effects become relevant. The extended
  peaking structure around a pseudoscalar mass of 2 TeV originates from the two chargino thresholds that are not affected by corrections due to strong interactions. It should be noted that the same corrections are valid for the reverse process $\gamma\gamma\to A$ as well, which could be probed at a potential future high-energy $\gamma\gamma$-collider.

\section{Conclusions} \label{sc:conclusions}
We have calculated the full SUSY--QCD corrections to pseudoscalar
Higgs-boson production via gluon fusion $gg\to A$ within the MSSM at
hadron colliders. We implemented the virtual stop and sbottom sector at
the NLO level to be in line with the necessities for the corresponding
scalar Higgs-boson production cross sections via gluon fusion $gg\to
h,H$. We have analyzed pseudoscalar Higgs-boson production with respect
to the introduction of effective low-energy top and bottom Yukawa
couplings, i.e.~the couplings within the low-energy 2HDM after
integrating out the strongly interacting SUSY particles (stops, sbottoms
and gluinos). We found that the bulk of the NLO corrections can be
absorbed in these effective Yukawa couplings, while the SUSY-remainder
is of moderate size, being significant close or above virtual squark
thresholds. We have analyzed the corrections in the context of the
Adler-Bardeen theorem and found that this theorem is fulfilled in the
large SUSY-mass limit, if the observable is expressed in terms of
properly matched low-energy parameters, i.e.~top- and bottom-Yukawa
couplings. The analogous results have also been obtained for the related
rare pseudoscalar Higgs-boson decays $A\to gg, \gamma\gamma$ that,
however, only play a minor role in phenomenological analyses at hadron colliders. This work
completes the full NLO QCD calculation for pseudoscalar MSSM Higgs
production and decay into gluonic and photonic final states and thus
serves as a basis for the corresponding theoretical predictions. \\

\noindent
{\bf Acknowledgements.} \\
We are grateful to J.~Reuter for private communication on the
Adler--Bardeen theorem. The research of T.T.D.N.~and M.M.~was supported by the Deutsche
Forschungsgemeinschaft (DFG, German Research Foundation) under grant
396021762 - TRR 257. The work of L.F.~has been supported by the Swiss National Science Foundation (SNSF).

\end{document}